\documentclass[11pt]{article}
\usepackage{amssymb}
\usepackage{graphics}
\usepackage{epsfig}
\usepackage{a4wide}
\usepackage{slashed}

\begin{document}

\newcommand{\ppvq}{$pp \to W_H(Z_H) q_- + X~$ }
\newcommand{\ppwz}{$pp \to W_H^{\pm}  Z_H + X~$}
\newcommand{\qqwz}{$qq^{\prime} \to W_H^{+} Z_H~$}
\newcommand{\udwz}{$u\bar d \to W_H^+ Z_H~$}
\newcommand{\qqwzg}{$qq^{\prime} \to W_H^{+} Z_H +g~$}
\newcommand{\qgwzq}{$qg \to W_H^{+} Z_H + q^{\prime}~$}
\newcommand{\ppfinala}{$pp \to W_H^{\pm} Z_H \to W^{\pm} A_H H A_H+X $}
\newcommand{\ppfinalap}{$pp \to W_H^{+} Z_H \to W^{+} A_H H A_H+X $}
\newcommand{\ppfinalam}{$pp \to W_H^{-} Z_H \to W^{-} A_H H A_H+X $}
\newcommand{\ppfinalb}{$pp \to W_H^{\pm} Z_H \to W^{\pm} A_H H A_H \to \mu^{\pm}\stackrel{(-)}{\nu_{\mu}} A_H H A_H+X$}
\newcommand{\ppfinalbp}{$pp \to W_H^{+} Z_H \to \mu^{+}{\nu_{\mu}} A_H H A_H+X$}
\newcommand{\ppfinalbm}{$pp \to W_H^{-} Z_H \to \mu^{-}\bar \nu_{\mu} A_H H A_H+X$}

\title{ Precise QCD predictions on $W_HZ_H$ production in the littlest Higgs Model with $T$ parity at the LHC    }
\author{ Liu Wen, Zhang Ren-You, Guo Lei, Ma Wen-Gan and Chen Liang-Wen  \\
{\small  Department of Modern Physics, University of Science and Technology}  \\
{\small  of China (USTC), Hefei, Anhui 230026, P.R.China}}

\date{}
\maketitle \vskip 15mm
\begin{abstract}
We investigate the effects of the littlest Higgs model with $T$
parity up to the QCD next-to-leading order (NLO) on the $W_H^{\pm}
Z_H$ productions at the CERN Large Hadron Collider (LHC), and
discuss the kinematic distributions of final decay products and the
theoretical dependence of the cross section on the
factorization/renormalization scale. We find the QCD NLO corrections
reduce the scale uncertainty of the leading order cross section in
case of $\mu_F=\mu_R$. By adopting the PROSPINO subtraction scheme
(scheme (II)) in analysing the QCD NLO contributions, we can obtain
the numerical results which keep the convergence of the perturbative
QCD description. Our results by adopting scheme (II) at the $14~{\rm
TeV}$ ($8~{\rm TeV}$) LHC show that the $K$-factor for the $W_H^+
Z_H$ production varies in the range of $1.01 \sim 1.10$ ($1.00 \sim
1.08$), while the $K$-factor for the $W_H^- Z_H$ production varies
in the range of $1.11 \sim 1.13$ ($1.11 \sim 1.12$), when the global
symmetry breaking scale $f$ goes from $400~ {\rm GeV}$ to $1.5~ {\rm
TeV}$ ($1~ {\rm TeV}$).
\end{abstract}


\vfill \eject \baselineskip=0.32in

\renewcommand{\theequation}{\arabic{section}.\arabic{equation}}
\renewcommand{\thesection}{\Roman{section}.}
\newcommand{\nb}{\nonumber}

\newcommand{\Dir}{\kern -6.4pt\Big{/}}
\newcommand{\Dirin}{\kern -10.4pt\Big{/}\kern 4.4pt}
\newcommand{\DDir}{\kern -7.6pt\Big{/}}
\newcommand{\DGir}{\kern -6.0pt\Big{/}}

\makeatletter      
\@addtoreset{equation}{section}
\makeatother       

\par
\section{Introduction }
\par
To interpret the mechanism of electroweak symmetry breaking
and resolve the little hierarchy problem \cite{LHP} are the major
motivations for the little Higgs models \cite{LittleHiggs}. In those
models some new gauge bosons, scalars and fermions are introduced at
a global symmetry breaking scale $f$ to cancel the one-loop
quadratic divergences for the Higgs mass from the standard model
(SM) \cite{s1,s2} particles. It deserves much attention due to their
elegant solution to the hierarchy problem and they are proposed as
one kind of electroweak symmetry breaking models without 
fine-tuning. Among the little Higgs
models there is one simplest version, the littlest Higgs (LH) model,
providing a set of new heavy gauge bosons ($W_H, Z_H, A_H$) and a
vector-like quark ($T$) to implement the divergence cancellation.
Nevertheless, precision electroweak measurements \cite{Limit-f}
severely constrain the LH model, especially the recent experimental
measurements \cite{Wmass,Zmass} on the searching for $W_H$ and $Z_H$
bosons.

\par
The precision electroweak constraints require the LH model
characterize a large value of $f$. To avoid fine-tuning between the
global symmetry breaking scale $f$ and the electroweak symmetry
breaking scale, a discrete symmetry named $T$ parity
\cite{Low:2004xc}-\cite{Hubisz:2005tx} is imposed. In this way, the
heavy gauge bosons assigned to be $T$-odd particles do not directly
couple with a pair of SM fermions and all dangerous tree-level
contributions to the precision measurements are forbidden,
therefore, the phenomenological constraints are somewhat relaxed.
Thus the LH model with $T$ parity (LHT)
\cite{Low:2004xc}-\cite{Cheng:2003ju} deserves more attention. In
the LHT, heavy gauge bosons, heavy fermions and heavy leptons
acquire masses through the breaking of the global symmetry, and
there exists an attractive dark matter candidate $A_H$ \cite{Asano}.
The global symmetry breaking scale $f$ can be lower than $1~{\rm
TeV}$ \cite{Hubisz:2005tx}, and the processes $W_H^{\mp}\to
l^{\mp}\stackrel{(-)}{\nu}$ and $Z_H\to l^{+}l^{-}$ are forbidden
due to the $T$ parity conservation, leaving the only $T$-odd heavy
gauge boson decay modes $W_H^{\pm} \to A_H W^{\pm}$ and $Z_H\to A_H
H$, where $H$ is the lightest neutral Higgs boson, followed by the
subsequential leptonic decays of $W^{\pm}$ and Higgs boson. As a
result, the experimental constraints \cite{Wmass,Zmass} on $m_{W_H}$
and $m_{Z_H}$ can not be applied to the $T$-odd gauge bosons in the
LHT. Recently, some QCD NLO phenomenological aspects of the LHT have
been analyzed in Refs.\cite{YanH, DuSM}. The $W_H Z_H$ production at
the LHC can be significant in searching for the new gauge bosons due
to the potential of its copious productions as shown in
Refs.\cite{cpyuan:2006ph,qhcao}, where the $W_H Z_H$ production at
the LHC is only studied at the leading-order (LO).

\par
The purpose of this work is to perform a comprehensive analysis for the
processes \ppwz at the LHC up to the QCD NLO. In Sec.II a
brief review of the related LHT theory is given. In Sec.III we present
the details of the calculations. The numerical results and
discussions are provided in Sec.IV. Finally we give a short summary.

\vskip 5mm
\section{Related LHT theory }\label{theory}
\par
In order to fix notations used in this paper we briefly review the
relevant LHT theory. The details of the LHT theory can be found in
Refs.\cite{Low:2004xc, Hubisz:2004ft,Hubisz:2005tx,cpyuan:2006ph}.

\par
In the LHT the assumed global symmetry $SU(5)$ is broken down
spontaneously to $SO(5)$ at some high scale $f$ around $1~{\rm TeV}$
\cite{LH5}. Breaking of $SU(5)$ leads to 14 massless
Nambu-Goldstone bosons, which transform under the electroweak gauge
group, $SU(2)_L \times U(1)_Y$, as a real singlet, a real triplet, a
complex doublet and a complex triplet. Four of the Nambu-Goldstone
bosons are treated as longitudinal components of the heavy gauge
bosons. The others decompose into a $T$-even $SU(2)$ doublet $h$,
identified as the SM Higgs doublet, and a complex $T$-odd $SU(2)$
triplet $\Phi$.

\par
The $T$ parity transformations for the gauge sector are defined as
the exchange between the gauge bosons of the two $SU(2)\times U(1)$
groups, i.e., $W_1^a \leftrightarrow W_2^a$ and $B_1 \leftrightarrow
B_2$. The gauge couplings of the two gauge groups have to be equal,
i.e., $g_1=g_2=\sqrt{2}g$ and $g^{\prime}_1=g^{\prime}_2=\sqrt{2}
g^{\prime}$. Thus their $T$-odd and $T$-even combinations can be
obtained as
\begin{eqnarray}
W_H^a=\frac{1}{\sqrt{2}}(W_1^a-W_2^a),&
B_H=\frac{1}{\sqrt{2}}(B_1-B_2),& (T-{\rm odd}), \nb\\
W_L^a=\frac{1}{\sqrt{2}}(W_1^a+W_2^a),&
B_L=\frac{1}{\sqrt{2}}(B_1+B_2),& (T-{\rm even}).
\end{eqnarray}
The mass eigenstates of the gauge sector in the LHT are expressed as
\begin{eqnarray}
W_H^{\pm}=\frac{1}{\sqrt 2}(W_H^1\mp i W_H^2), &
Z_H=\sin{\theta_{H}} B_H+ \cos{\theta_{H}} W^3_H,&
A_H=\cos{\theta_{H}} B_H-\sin{\theta_{H}} W^3_H , \nb \\
W_L^{\pm}=\frac{1}{\sqrt 2}(W_L^1\mp i W_L^2), &
Z_L=-\sin{\theta_{W}} B_L+ \cos{\theta_{W}} W^3_L,&
A_L=\cos{\theta_{W}} B_L+\sin{\theta_W} W^3_L,
\end{eqnarray}
where $\theta_W$ is the Weinberg angle, and the mixing angle
$\theta_H$ at the ${\cal O}(v_{SM}^2/f^2)$ is expressed as
\begin{eqnarray}
\sin \theta_H \simeq \left[ \frac{5gg'}{4(5g^2-g'^2)}\frac{v_{SM}
^2}{f^2} \right].
\end{eqnarray}
The $T$-even gauge bosons $A_L$, $Z_L$ and $W_L^{\pm}$ are
identified with the SM gauge bosons, while the four new heavy gauge
bosons, the $T$-odd partners of SM gauge bosons, are $A_H$, $Z_H$
and $W_H^{\pm}$ with masses of \cite{cpyuan:2006ph}
\begin{eqnarray} \label{m_v}
m_{A_H}\simeq
\frac{1}{\sqrt{5}}g'f\left(1-\frac{5}{8}\frac{v_{SM}^2}{f^2}\right),
& m_{Z_H}\simeq m_{W_H} \simeq
gf\left(1-\frac{1}{8}\frac{v_{SM}^2}{f^2}\right),
\end{eqnarray}
where $v_{SM}= 246~ {\rm GeV}$. The $T$ parity partner of photon,
$A_H$, is the lightest $T$-odd particle. Therefore, the heavy photon
is a candidate of dark matter. The masses of SM gauge bosons can be
expressed as $m_W = \frac{1}{2} g v_{SM}$ and $m_Z = \frac{1}{2}
\sqrt{g^2 + g^{\prime 2}} v_{SM}$ at the tree-level.

\par
When the $T$ parity is implemented in the fermion sector of the
model, the existence of mirror partners for each of the original
fermions are required. The $T$-odd partners of SM up- and down-type
quarks are denoted as $U_{-}$ and $D_{-}$, where $U_{-}=u_-,c_-,t_-$
and $D_{-}=d_-,s_-,b_-$. We can get their masses as
\cite{cpyuan:2006ph}
\begin{eqnarray} \label{m_q}
m_{U_{-}}\simeq \sqrt{2}\kappa
f\left(1-\frac{1}{8}\frac{v_{SM}^2}{f^2}\right), &&
m_{D_{-}}=\sqrt{2}\kappa f,
\end{eqnarray}
where $\kappa$ is the mass coefficient in Lagrangian of the quark
sector. The Feynman rules in the LHT used in this work are presented
in Appendix.

\vskip 5mm
\par
\section{Calculations }\label{calc}
\par
In the LO and QCD NLO calculations we employ the FeynArts 3.4
package \cite{FeynArts} to generate Feynman diagrams and their
corresponding amplitudes. To implement the amplitude calculations we
apply FormCalc 5.4 programs \cite{FormCalc}. The t'Hooft-Feynman
gauge and the five-flavor scheme (5FS) are adopted in this work.

\par
\subsection{LO cross section }
\par
\begin{figure}
\begin{center}
\includegraphics[width=0.6\textwidth]{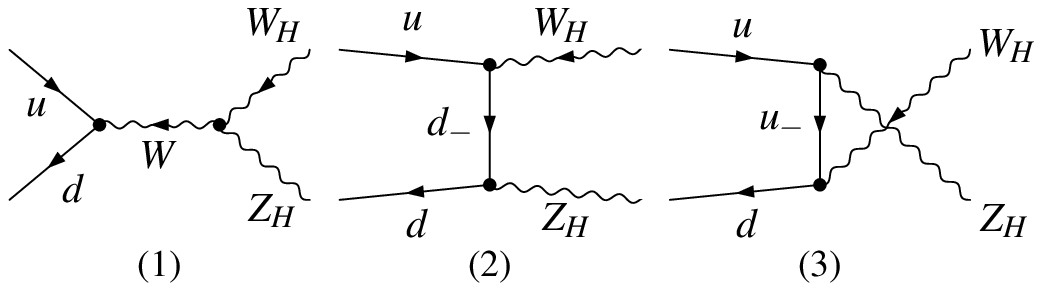}
\caption{ \label{fig1} The LO Feynman diagrams for the partonic
process \udwz. }
\end{center}
\end{figure}

At the parton level the cross section for the $qq^{\prime}  \to
W_H^-Z_H$ $(q q^{\prime}= \bar u d, \bar u s, \bar c d, \bar c s)$
subprocess in the LHT should be the same as that for the
corresponding charge conjugate subprocess $qq^{\prime} \to W_H^+Z_H$
$(q q^{\prime}= u \bar d, u \bar s, c \bar d, c \bar s)$ due to the
$CP$-conservation. We present the parton level calculations for the
related subprocess $qq^{\prime} \to W_H^+Z_H$ in this section. By
neglecting the contribution of bottom quark in the initial state,
the LO contribution to the cross section for the parent process $pp
\to W_H^{+} Z_H+X$ comes from the subprocesses
\begin{eqnarray}
\label{process} q(p_{1})+q^{\prime}(p_{2})\to
W_H^{+}(p_{3})+Z_H(p_{4}), && (q q^{\prime}= u \bar d, u \bar s, c
\bar d, c \bar s),
\end{eqnarray}
where $p_i~(i = 1, 2, 3, 4)$ represent the four-momenta of the
incoming partons and the outgoing $W_H^{+}$, $Z_H$ bosons,
respectively. The Feynman diagrams for the $u\bar{d} \to W_H^+ Z_H$
partonic process are shown in Fig.\ref{fig1}, and the LO Feynman
graphs for other relevant partonic processes $q q^{\prime} \to
W_H^{+} Z_H$ ($q q^{\prime}= u \bar s, c \bar d, c \bar s$) are
similar with those in Fig.\ref{fig1}.

\par
The expression for the LO cross section for the partonic process
$qq^{\prime} \to W_H^{+} Z_H$ ($q q^{\prime}= u \bar d, u \bar s, c
\bar d, c \bar s$) has the form as
\begin{eqnarray}
\hat{\sigma}^0_{q q^{\prime}}= \frac{1}{4} \frac{1}{9}
\frac{1}{4|\vec{p}|\sqrt{\hat{s}}}\int \sum_{spin}\sum_{color}
|{\cal M}^{LO}_{q q^{\prime}}|^2 d\Omega_2,&& (q q^{\prime}= u \bar
d, u \bar s, c \bar d, c \bar s),
\end{eqnarray}
where the factors $\frac{1}{4}$ and $\frac{1}{9}$ come from
averaging over the spins and colors of the initial partons,
respectively, $\vec{p}$ is the three-momentum of one initial parton
in center-of-mass system, $\sqrt{\hat{s}}$ is the partonic center-of-mass system
energy and ${\cal M}_{q q^{\prime}}^{LO}$ is the amplitude of all
the tree-level diagrams for the partonic process $q q^{\prime} \to
W_H^+Z_H$. The summation is taken over the spins and colors of all
the relevant particles in the $q q^{\prime}\to W_H^{+} Z_H$
subprocess. We perform the integration over the two-body phase space
of the final particles $W_H^{+}$ and $Z_H$. The phase space element
$d\Omega_2$ is expressed as
\begin{eqnarray}
d\Omega_2 =(2 \pi )^4 \delta^{(4)}(p_1+p_2-p_3-p_4) \frac{d^3
\vec{p}_3}{(2\pi)^3 2E_3} \frac{d^3 \vec{p}_4}{(2\pi)^3 2E_4}.
\end{eqnarray}
Then the LO total cross section for the parent process $pp \to
W_H^{+} Z_H +X$ can be expressed as
\begin{eqnarray} \label{pp-total cross section}\sigma_{LO}
=\sum_{q q^{\prime}=u \bar d, u \bar s,} ^{c \bar d, c \bar s}
\int_0^1 dx_1 \int_0^1 dx_2 \left[
G_{q/P_1}(x_1,\mu_F)G_{q^{\prime}/P_2} (x_2,\mu_F)+(1
\leftrightarrow 2) \right] \hat{\sigma}^0_{q q^{\prime}}(\hat s =
x_1 x_2 s),
\end{eqnarray}
where $G_{j/P}$ $(j=u,c,\bar d,\bar s)$ is the parton distribution function
(PDF) of proton $P$,
which describes the probability in finding a parton $j$ with
momentum $xp_j$ in proton $P$, $s$ represents the total colliding
energy squared in the rest frame of proton-proton system, and
$\mu_F$ is the factorization scale.

\par
\subsection{QCD NLO corrections  }
\par
The QCD NLO corrections to the parent process $pp \to W_H^{+} Z_H +
X$ at the LHC can be divided into four parts:
\begin{itemize}
\item The QCD one-loop virtual corrections to the partonic processes
$qq^{\prime} \to W_H^{+} Z_H$;
\item The contributions of the real gluon emission partonic processes
$qq^{\prime} \to W_H^{+} Z_H + g$;
\item The contributions of the real light-(anti)quark emission partonic
processes $qg \to W_H^{+} Z_H + q^{\prime}$;
\item The contributions of the PDF counterterms.
\end{itemize}
The dimensional regularization method in $D=4-2\epsilon$
dimensions is adopted in this work to isolate the ultraviolet (UV)
and infrared (IR) singularities in the NLO calculations.

\par
\subsubsection{One-loop virtual corrections to \qqwz partonic process }
\par
\begin{figure}
\begin{center}
\includegraphics[width=0.82\textwidth]{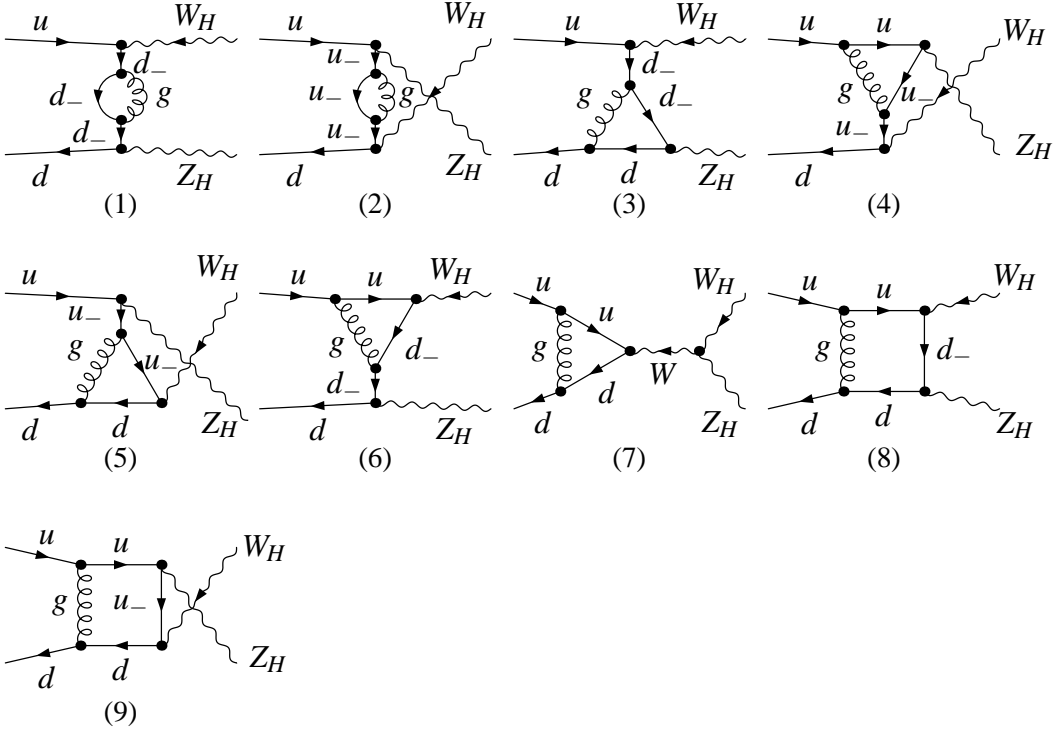}
\caption{ \label{fig2a} The representative one-loop Feynman diagrams
for the partonic process \udwz. }
\end{center}
\end{figure}

Some representative Feynman diagrams for the one-loop virtual corrections
to the partonic process \udwz are presented in Fig.\ref{fig2a}.
There exist both UV and IR singularities. The masses and wave functions
of SM quarks and their $T$-odd partners should be renormalized to remove
the UV divergences. The counterterms are defined as
\begin{eqnarray}
\psi^{0,L,R}_{q} & = & \left(1+\frac{1}{2}\delta Z_{q}^{L,R}\right)\psi^{L,R}_{q}~, \\
\psi^{0,L,R}_{q_-} & = & \left(1+\frac{1}{2}\delta Z_{q_-}^{L,R}\right)\psi^{L,R}_{q_-}~, \\
m^0_{q_-} & = & m_{q_-}+\delta m_{q_-}~,
\end{eqnarray}
where $\psi^{L,R}_{q},\psi^{L,R}_{q_-}$ denote the fields of SM
quark and $T$-odd quark, respectively, and $m_{q_-}$denotes the mass
of $T$-odd quark. The on-shell scheme is applied to renormalize the
relevant fields and masses, then we obtain
\begin{eqnarray}
\delta Z^{L,R}_{q} & = & - \frac{\alpha_s (\mu_R)}{3\pi} \left[
\Delta_{UV} -
\Delta_{IR} \right]~, \\
\delta Z^{L,R}_{q_-} & = & - \frac{\alpha_s (\mu_R)}{3\pi} \left[
\Delta_{UV} + 2\Delta_{IR} + 4 + 3\ln
\left(\frac{\mu_R^2}{m_{q_-}^2}
\right) \right]~, \\
\frac{\delta m_{q_-}}{m_{q_-}} & = & - \frac{\alpha_s(\mu_R)}{3\pi}
\left\{3\left[\Delta_{UV}+\ln\left(\frac{\mu_R^2}{m_{q_-}^2}\right)\right]+4
\right\}~,
\end{eqnarray}
where $\Delta_{UV}=\frac{1}{\epsilon_{UV}}-\gamma_E + \ln (4\pi)$
and $\Delta_{IR}=\frac{1}{\epsilon_{IR}}-\gamma_E + \ln (4\pi)$. The
one-loop virtual contribution is UV finite after performing the
renormalization procedure. Nevertheless, there still exist soft and
collinear IR singularities. By adding the contributions of the real
gluon/light-(anti)quark emission subprocesses and the counterterms
of the PDFs at the NLO, the remaining singularities are canceled as
we shall see later.

\par
\subsubsection{Real gluon/light-(anti)quark emission corrections}
\par
\begin{figure}
\begin{center}
\includegraphics[width=0.75\textwidth]{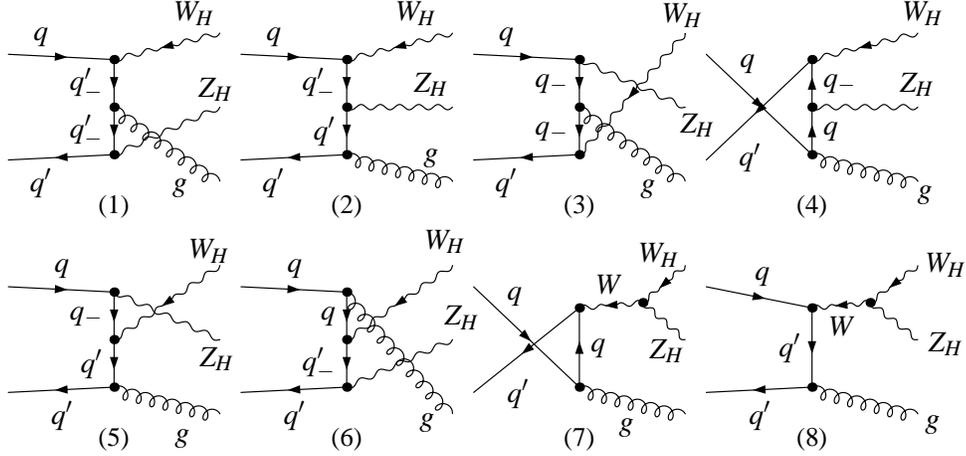}
\caption{ \label{fig2c} The tree-level Feynman diagrams for the real
gluon emission partonic process $qq^{\prime} \to W_H^+ Z_H +g$,
$(q q^{\prime}= u \bar d, u \bar s, c \bar d, c \bar s)$. }
\end{center}
\end{figure}

The real gluon emission partonic processes for the $W_H^{+}Z_H$
production can be denoted as
\begin{eqnarray}
q(p_{1})+ q^{\prime}(p_{2})\to W_H^{+}(p_{3})+Z_H(p_{4})+g(p_{5}) ,
&& (q q^{\prime}= u \bar d, u \bar s, c \bar d, c \bar s).
\end{eqnarray}
The real gluon emission subprocess $q q^{\prime} \to W_H^{+} Z_H g$
contains both soft and collinear IR singularities which can be
conveniently isolated by adopting the two cutoff phase space slicing
(TCPSS) method \cite{tcpss}. In Fig.\ref{fig2c} we show the tree
level Feynman diagrams for this subprocess. In performing the
calculations with the TCPSS method, we should introduce two
arbitrary small soft cutoff $\delta_s$ and collinear cutoff
$\delta_c$. The phase space of the $qq^{\prime} \to W_H^{+} Z_H g$
subprocess can be split into two regions: soft gluon region
($E_5\leq \frac{1}{2}\delta_s \sqrt{\hat s}$) and hard gluon region
($E_5> \frac{1}{2}\delta_s \sqrt{\hat s}$) by the soft cutoff
$\delta_s$. The hard gluon region is separated as hard collinear
($HC$) and hard noncollinear ($\overline{HC}$) regions by the
collinear cutoff $\delta_c$. The $HC$ region is the phase space
where $\hat s_{15}\leq \delta_c \hat s$ or $\hat s_{25}\leq \delta_c
\hat s$ ($\hat{s}_{ij}=(p_i+p_j)^2$). Then the cross section for the
real gluon emission subprocess is written as
\begin{eqnarray}
\hat{\sigma}^R_g =\hat{\sigma}^S_g+\hat{\sigma}^H_g=
\hat{\sigma}^S_g+\hat{\sigma}^{\rm HC}_g+\hat{\sigma}^{\overline{\rm
HC}}_g.
\end{eqnarray}
According to the Kinoshita-Lee-Nauenberg (KLN) theorem \cite{KLN},
the soft singularity in the soft part $\hat{\sigma}_g^{S}$ can be
canceled by the soft IR divergence in the virtual corrections, while
the hard noncollinear cross section part
$\hat{\sigma}_g^{\overline{HC}}$ is IR safe. The virtual corrections
cancel part of the collinear singularity and the PDF
counterterms absorb the remaining collinear divergence.

\par
\begin{figure}
\begin{center}
\includegraphics[width=0.75\textwidth]{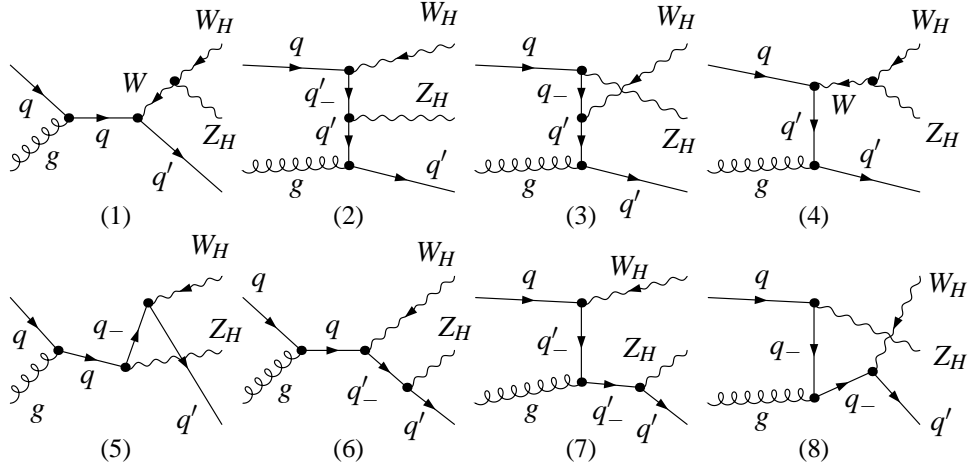}
\caption{ \label{fig3} The tree-level Feynman diagrams for the real
light-quark emission partonic process $qg \to W_H^+ Z_H +q^{\prime}$
$(qq^{\prime}= ud, cs, \bar{d}\bar{u}, \bar{s}\bar{c})$.}
\end{center}
\end{figure}

Beside the real gluon emission subprocesses, the real
light-(anti)quark emission subprocesses, which have the same order
contributions with the real gluon emission subprocesses, should
be taken into account. This kind of subprocesses is denoted as
\begin{eqnarray}
q(p_{1})+ g(p_{2})\to W_H^{+}(p_{3})+Z_H(p_{4})+q^{\prime}(p_{5}) ,
&& (qq^{\prime}= ud, cs, \bar{d}\bar{u}, \bar{s}\bar{c}).
\end{eqnarray}
The corresponding Feynman diagrams for the subprocess $qg \to W_H^+
Z_H +q^{\prime}$ at the tree-level are shown in Fig.\ref{fig3}.
Using the TCPSS method described above, the phase space can be split
into a collinear ($C$) region ($\hat s_{15}\leq \delta_c \hat s$ or
$\hat s_{25}\leq \delta_c \hat s$) and a noncollinear
($\overline{C}$) region ($\hat s_{15} > \delta_c \hat s$ and $\hat
s_{25} > \delta_c \hat s$) by a collinear cutoff $\delta_c$.
Therefore, the cross section for the real light-(anti)quark emission
subprocess can be expressed as
\begin{eqnarray}
\hat{\sigma}^R_{q}= \hat{\sigma}_{q}^{C}+
\hat{\sigma}_{q}^{\overline{C}}.
\end{eqnarray}
The cross section $\hat{\sigma}_{q}^{\overline{C}}$ in the
noncollinear region is finite and can be evaluated in
four dimensions using Monte Carlo technique while
$\hat{\sigma}_{q}^{C}$ contains collinear singularity. After adding
the renormalized virtual corrections and the real
gluon/light-(anti)quark emission corrections to the subprocess
\qqwz, the partonic cross section still contains the collinear
divergence, which can be absorbed into the redefinition of the PDFs
at the NLO.

\par
\subsubsection{PDF counterterms}
\par
The PDF counterterms, $\delta G_{i/P}(x,\mu_F)$ ($i=g,u,\bar
u,d,\bar d,c,\bar c, s,\bar s$), which absorb the remaining
collinear divergence, can be split into two parts: the collinear
gluon emission part $\delta G_{i/P}^{(gluon)}(x,\mu_F)$ and the
collinear light-quark emission part $\delta
G_{i/P}^{(quark)}(x,\mu_F)$:
\begin{eqnarray} \label{PDFcounterterm1}
&& \delta G_{q(g)/P}(x,\mu_F) = \delta G_{q(g)/P}^{(gluon)}(x,\mu_F)
                  +\delta G_{q(g)/P}^{(quark)}(x,\mu_F),
~~~(q = u, \bar{u}, d, \bar{d}, c, \bar{c}, s, \bar{s} ),
\end{eqnarray}
where
\begin{eqnarray} \label{PDFcounterterm2}
&& \delta G_{q(g)/P}^{(gluon)}(x,\mu_F) =
   \frac{1}{\epsilon} \left[
                      \frac{\alpha_s}{2 \pi}
                      \frac{\Gamma(1 - \epsilon)}{\Gamma(1 - 2 \epsilon)}
                      \left( \frac{4 \pi \mu_R^2}{\mu_F^2} \right)^{\epsilon}
                      \right]
   \int_x^1 \frac{dz}{z} P_{qq(gg)}(z) G_{q(g)/P}(x/z,\mu_F), \nonumber \\
&& \delta G_{q/P}^{(quark)}(x,\mu_F) =
   \frac{1}{\epsilon} \left[
                      \frac{\alpha_s}{2 \pi}
                      \frac{\Gamma(1 - \epsilon)}{\Gamma(1 - 2 \epsilon)}
                      \left( \frac{4 \pi \mu_R^2}{\mu_F^2} \right)^{\epsilon}
                      \right]
   \int_x^1 \frac{dz}{z} P_{qg}(z) G_{g/P}(x/z,\mu_F),  \nonumber \\
&& \delta G_{g/P}^{(quark)}(x,\mu_F) =
   \frac{1}{\epsilon} \left[
                      \frac{\alpha_s}{2 \pi}
                      \frac{\Gamma(1 - \epsilon)}{\Gamma(1 - 2 \epsilon)}
                      \left( \frac{4 \pi \mu_R^2}{\mu_F^2} \right)^{\epsilon}
                      \right]
   \sum_{q=u,\bar{u},d,\bar{d},}^{c, \bar {c}, s, \bar {s},b,\bar{b},}
   \int_x^1 \frac{dz}{z} P_{gq}(z) G_{q/P}(x/z,\mu_F).~~~
\end{eqnarray}
More details about the explicit expressions for the splitting functions
$P_{ij}(z) (ij=qq,qg,gq,gg)$ are available in Ref.\cite{tcpss}.

\par
\subsubsection{Total QCD NLO correction }
\par
Finally, we have eliminated all the UV and IR singularities by
performing the renormalization procedure and adding all the QCD NLO
correction components, and we get the finite QCD NLO corrected
integrated cross section for the $pp \to W_H^{+}Z_H+X$ process as
\begin{eqnarray}\label{TotalCorr}
\sigma_{NLO}&=&\sigma_{LO}+\Delta
\sigma_{NLO}= \sigma_{LO}+\Delta\sigma^{(2)}+\Delta\sigma^{(3)}.
\end{eqnarray}
The two-body term $\Delta \sigma^{(2)}$ includes the one-loop
corrections to the $pp \to W_H^{+}Z_H+X$ process and the tree-level
contributions in the soft and hard collinear regions for the real
gluon/light-(anti)quark emission processes, while the three-body
term $\Delta \sigma^{(3)}$ contains the cross sections for the real
gluon/light-(anti)quark emission processes over the hard noncollinear region.

\par
In this work, two event selection schemes are applied in discussing
the QCD NLO corrections. In scheme (I) all the NLO correction
components mentioned above are included in the QCD NLO corrections,
called also the inclusive event selection scheme.
In this scheme, there exists resonance effect in Figs.\ref{fig3}
(5)-(8) duo to the possible on-shell $q_-$ propagator and those Feynman
diagrams could lead to large corrections to the Born $pp \to
W_H^{+}Z_H + X$ process, so that the perturbative convergence
would be eventually destroyed. To deal
with the resonance effect in these partonic processes, the $q_-$
mass squared $m_{q_-}^2$ in its propagator should be replaced by
$m_{q_-}^2-im_{q_-}\Gamma_{q_-}$. The partial decay widths of
$T$-odd quarks are obtained numerically by adopting the expressions
presented in Ref.\cite{DuSM}.

\par
Actually, the contributions from the diagrams for the $qg \to W_H^+
Z_H +q^{\prime}$ subprocess with intermediate on-shell $T$-odd quark
$q_-$ shown in Figs.\ref{fig3} (5)-(8), should pertain to other
on-shell $W_H q_-$ and $Z_H q_-$ associated production channels,
i.e., $pp \to qg \to W_H q_-^{\prime}+X$  and $pp \to qg \to Z_H
q_-+X$ processes, followed with subsequential decays of
$q_-^{\prime} \to Z_H q^{\prime}$ and $q_- \to W_H q^{\prime}$,
respectively. To avoid double counting and to keep the convergence
of the perturbative QCD description for the $pp \to W_H^+ Z_H+X$
process, we adopt the PROSPINO subtraction strategy \cite{PROSPINO,
on-shell} to remove the on-shell $T$-odd quark $q_-$ contributions
called scheme (II). This subtraction scheme can provide a reliable
production rate since it only subtracts the squared on-shell
amplitudes and does this point by point over the entire phase space.
The PROSPINO subtraction is done by performing a replacement of the
Breit-Wigner propagator
\begin{eqnarray}
 \frac{|{\cal M}|^2( s_{V_H q} )}{( s_{V_H q} - m_{q_-}^2 )^2
 + m_{q_-}^2 \Gamma_{q_-}^2}
 & \to &
 \frac{|{\cal M}|^2( s_{V_H q} )}{( s_{V_H q} - m_{q_-}^2 )^2
 + m_{q_-}^2 \Gamma_{q_-}^2} \nonumber \\
 &&-
 \frac{|{\cal M}|^2( m_{q_-}^2 )}{( s_{V_H q} - m_{q_-}^2 )^2
 + m_{q_-}^2 \Gamma_{q_-}^2}
 \Theta( \hat{s} - 4 m_{q_-}^2 )
 \Theta( m_{q_-} - m_{V_H} ),
\end{eqnarray}
where $s_{V_H q}$ is the squared momentum flowing through the
intermediate $q_-$ propagator.

\par
Analogously, we can follow above calculation procedure to evaluate
the LO and NLO QCD corrected results for the $pp \to W_H^- Z_H+X$
process at the LHC.

\vskip 5mm
\section{Numerical results and discussions }\label{numres}
\par
\subsection{Input parameters}\label{parameters}
\par
The two mixing matrices, $V_{Hu}$ and $V_{Hd}$ cannot be set to be
unit matrices simultaneously due to the condition of
$V_{Hu}^{\dag}V_{Hd}=V_{CKM}$ \cite{Blanke:2007ckm}. In our
numerical calculations $V_{Hu}$ is set as a unit matrix, then we get
$V_{Hd}=V_{CKM}$. We take $\alpha_{{\rm ew}}(m_Z^2)^{-1}=127.916$,
$m_W=80.399~{\rm GeV}$, $m_Z=91.1876~{\rm GeV}$ and
$\sin^2\theta_W=1-\left(\frac{m_W}{m_Z}\right)^2=0.2226$
\cite{databook}. We neglect the masses of $\mu$-lepton and light
quarks. The colliding energy in the proton-proton center-of-mass
system is set as $\sqrt{s}=8~{\rm TeV}$ for the early LHC and
$\sqrt{s}=14~{\rm TeV}$ for the future LHC. We define
$\mu_0=(m_{W_H} +m_{Z_H})/2$ and adopt CTEQ6L1 and CTEQ6M PDFs in
the LO and NLO calculations, respectively. The LHT $T$-odd quark
mass coefficient parameter $\kappa$ is fixed to be 1. Consequently
the masses of heavy gauge bosons and $T$-odd quarks are only the
functions of the LHT parameter $f$ as shown in Eqs.(\ref{m_v}) and
(\ref{m_q}). The Cabibbo-Kobayashi-Maskawa (CKM) matrix elements are
taken as
\begin{eqnarray}\label{CKM}
 V_{CKM} &=& \left(
\begin{array}{ccc}
    V_{ud} \ &  V_{us} \ &  V_{ub} \\
    V_{cd} \ &  V_{cs} \ &  V_{cb} \\
    V_{td} \ &  V_{ts} \ &  V_{tb} \\
\end{array}
    \right)=\left(
\begin{array}{ccc}
     ~~0.97418 \ &  0.22577 \ &  0 \\
    -0.22577 \ &  0.97418 \ &  0 \\
       0 \ &  0 \ &  1 \\
\end{array}  \right).
\end{eqnarray}

\par
By using Eqs.(\ref{m_v}-\ref{m_q}) and taking the LHT parameter
$\kappa = 1$, we obtain the masses of heavy gauge bosons and $T$-odd
quarks for some typical values of the LHT global symmetry breaking
scale $f$ and list them in Table \ref{tab1}.
\begin{table}
\begin{center}
\begin{tabular}{c|c|c|c|c}
  \hline
    $f$    &  $m_{W_H}\approx m_{Z_H}$ & $m_{A_H}$  & $m_{u_-}=m_{c_-}$  & $m_{d_-}=
m_{s_-}$  \\
   $~~({\rm GeV})~~$ & $~~({\rm GeV})~~$ & $~~({\rm GeV})~~$  & $~~({\rm GeV})~~$ & $~~({\rm GeV})~~$ \\
  \hline
     500  & 322.1  & 67.5  & 685.7   & 707.1    \\
     700  & 457.8  & 102.7 & 974.7   & 989.9    \\
     800  & 525.1  & 119.7 & 1118.0  & 1131.4   \\
     900  & 592.3  & 136.4 & 1260.9  & 1272.8   \\
     1000 & 659.3  & 153.0 & 1403.5  & 1414.2   \\
     1100 & 726.1  & 169.4 & 1545.9  & 1555.6   \\
     1300 & 859.7  & 202.0 & 1830.3  & 1838.5   \\
     1500 & 993.1  & 234.5 & 2114.2  & 2121.3   \\
  \hline
\end{tabular}
\end{center}
\begin{center}
\begin{minipage}{15cm}
\caption{\label{tab1} The masses of $W_H$, $Z_H$, $A_H$ and $q_-$
($q_-=u_-,d_-,c_-,s_-$) for some typical values of the LHT parameter
$f$ with $\kappa =1$. }
\end{minipage}
\end{center}
\end{table}

\par
\subsection{Checks}
\par
The correctness of our calculations are verified through the
following aspects:

\par
{\bf 1.} Our LO cross sections are in good agreement with the results
read out from Fig.9 of Ref.\cite{cpyuan:2006ph} when we employ the
same input parameters and PDFs as used in Ref.\cite{cpyuan:2006ph}.

\par
{\bf 2.} After combining all the contributions at the QCD NLO, the
cancelations of UV and IR divergences are verified.

\par
\begin{figure}[htbp]
\begin{center}
\includegraphics[scale=0.75]{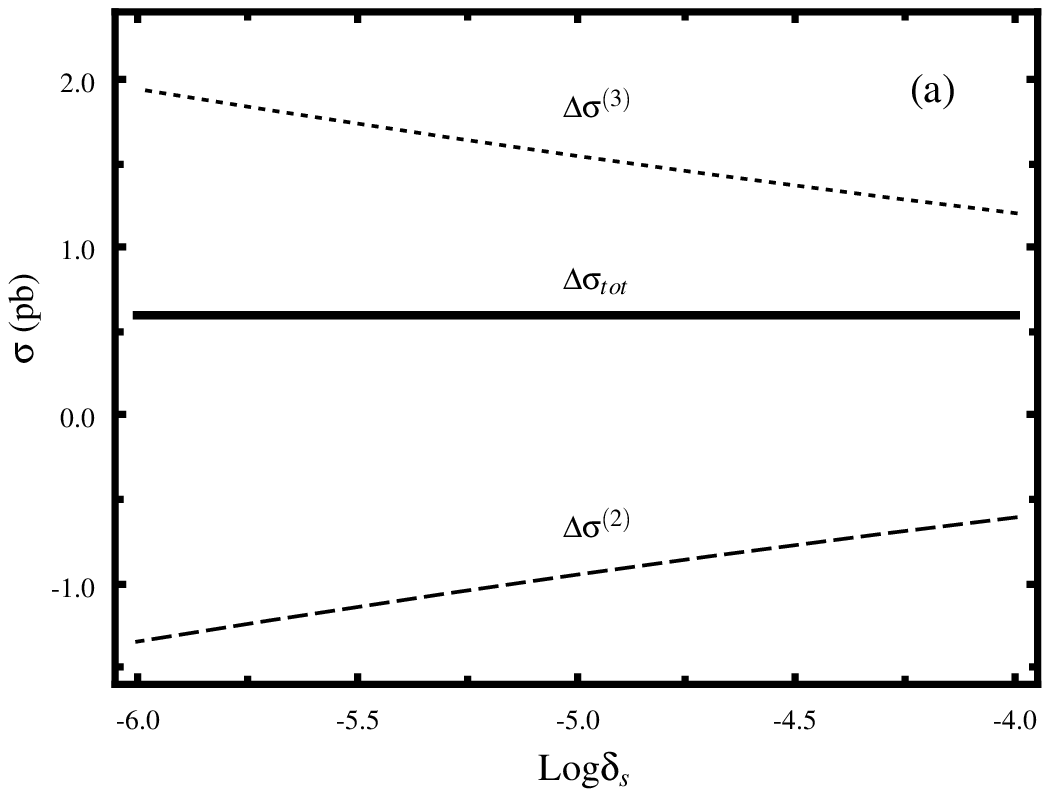}
\includegraphics[scale=0.75]{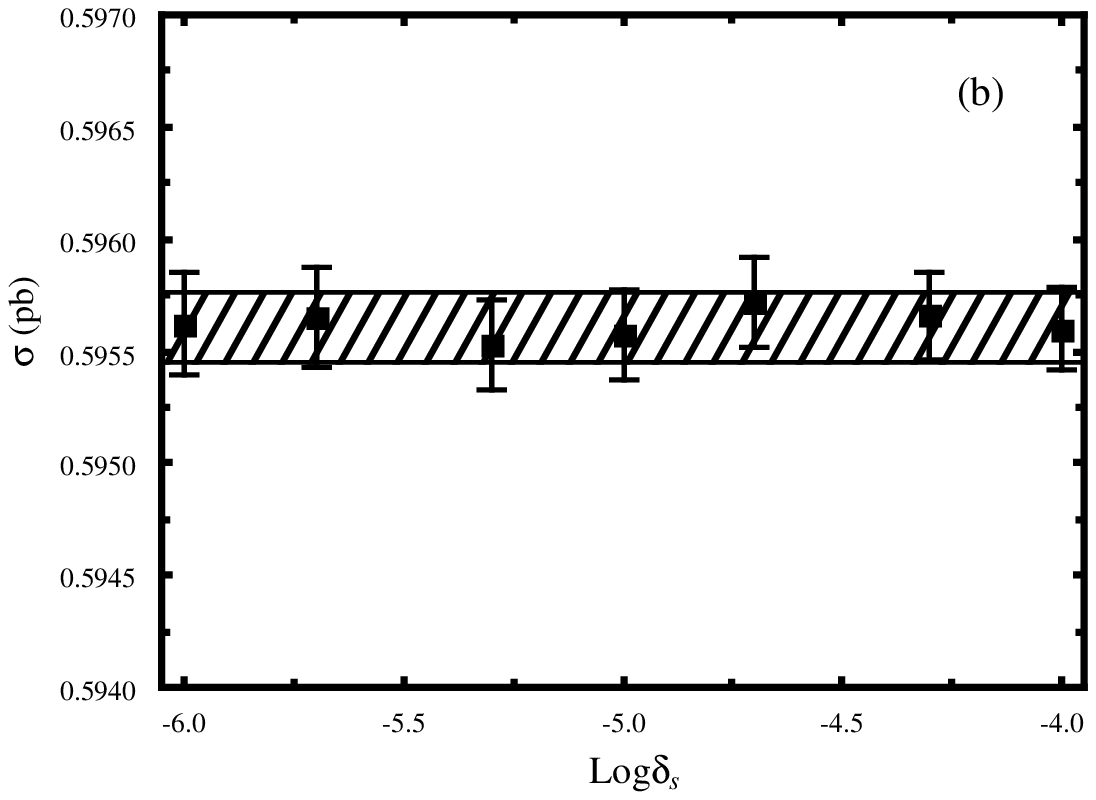}
\hspace{0in}%
\caption{\label{fig5} (a) The dependence of the QCD NLO corrections
to the $pp \to u\bar d \to W_H^+ Z_H + X$ process on the cutoffs
$\delta_s$ and $\delta_c$ at the $\sqrt{s}=14~{\rm TeV}$ LHC, where we
take $f=600~{\rm GeV}$, $\kappa=1$, $\delta_c=\delta_s/100$ and
$\mu=\mu_0=390.20~{\rm GeV}$. (b) The amplified curve for
$\Delta\sigma_{tot}$ in Fig.\ref{fig5}(a). The shadowing region
shows the result by adopting the DPS method with $\pm 1\sigma$
statistic error. }
\end{center}
\end{figure}

{\bf 3.} We make the verification of the $\delta_s/\delta_c$
independence of the total QCD NLO correction, where two arbitrary
cutoffs $\delta_s$ and $\delta_c$ \cite{tcpss} are introduced to
separate the phase space in order to isolate the soft and collinear
IR divergences, respectively. Eq.(\ref{TotalCorr}) shows that the
total QCD NLO correction ($\Delta \sigma_{NLO}$) is obtained by
summing up the two-body and three-body corrections ($\Delta
\sigma^{(2)}$ and $\Delta \sigma^{(3)}$). We depict $\Delta
\sigma^{(2)}$, $\Delta \sigma^{(3)}$ and $\Delta \sigma_{NLO}$ for
the process $pp \to u \bar{d} \to W_H^+ Z_H+X$ as functions of the
soft cutoff $\delta_s$ in Fig.\ref{fig5}(a) with $f=600~{\rm GeV}$,
$\kappa=1$, $\delta_c=\delta_s/100$ and $\mu=\mu_0=(m_{W_H} +
m_{Z_H})/2=390.20~{\rm GeV}$. The amplified curve for the total correction
$\Delta\sigma_{NLO}$ in Fig.\ref{fig5}(a) is demonstrated in
Fig.\ref{fig5}(b) together with calculation errors. From these two
figures we find that the total QCD NLO correction $\Delta
\sigma_{NLO}$ is independent of the two cutoffs within the
statistical errors. This independence is an indirect check for the
correctness of our work. We adopt also the dipole subtraction (DPS)
method \cite{dipole} to deal with the IR singularities. The total
QCD NLO correction $\Delta \sigma_{NLO}$ obtained by adopting the
DPS method with $\pm 1\sigma$ statistic error is plotted as the
shadowing region in Fig.\ref{fig5}(b). We can see that the results
from both the TCPSS method and the DPS method are in good agreement.
In further numerical calculations, we fix $\delta_s = 1\times
10^{-4}$ and $\delta_c=1\times 10^{-6}$.

\par
\subsection{Dependence on factorization/renormalization scale  }
\par
\begin{figure}[htbp]
\begin{center}
\includegraphics[width=0.60\textwidth]{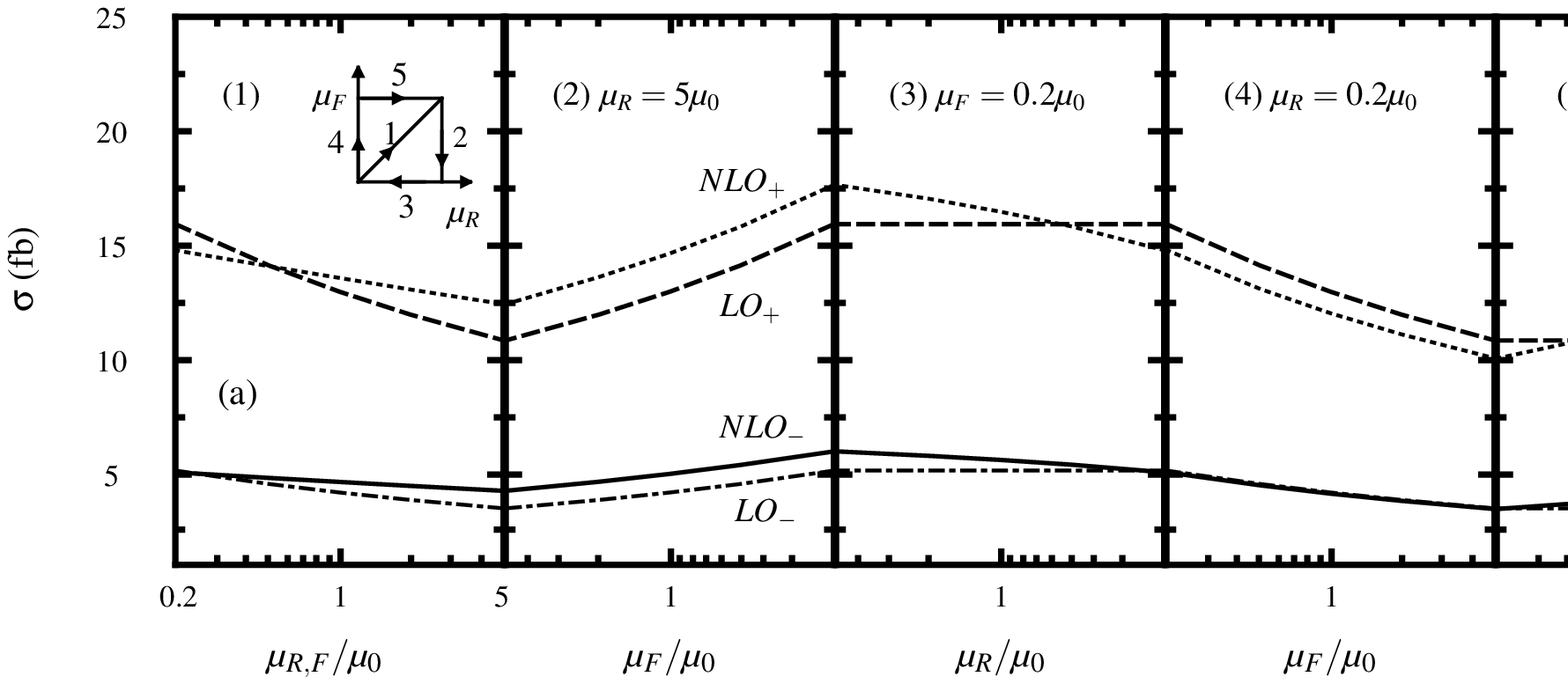}
\\~~\\
\includegraphics[width=0.60\textwidth]{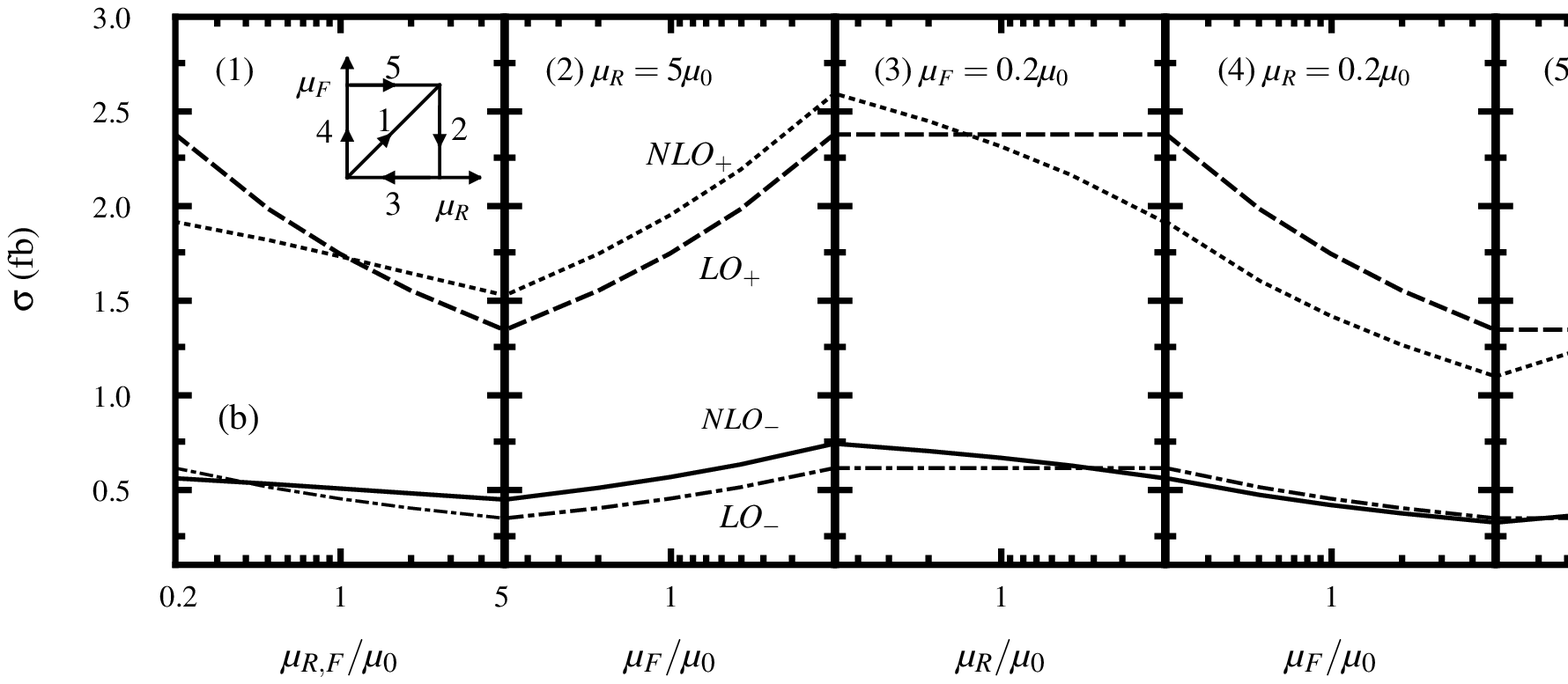}
\hspace{0in}%
\caption{\label{fig6} Profile of the renormalization and
factorization scale dependence of the LO and NLO corrected cross
sections for the processes $pp \to W_H^+Z_H+X$ and $pp \to
W_H^-Z_H+X$. The two plots trace the scale dependence following a
contour in the $\mu_R-\mu_F$ plane. There we take the (II) selection
scheme and assume $\mu/\mu_0=0.2-5$, the LHT parameters $f=1~{\rm
TeV}$ and $\kappa = 1$. (a) at the $\sqrt{s}=14~{\rm TeV}$ LHC. (b)
at the $\sqrt{s}=8~{\rm TeV}$ LHC. }
\end{center}
\end{figure}

In order to investigate whether the production rates for the $pp \to
W_H^+Z_H+X$ and $pp \to W_H^-Z_H+X$ processes at the
$\sqrt{s}=14~{\rm TeV}$ and the $\sqrt{s}=8~{\rm TeV}$ LHC have the
stabilization of the dependence on the unphysical renormalization
and the factorization scales, we present Figs.\ref{fig6}(a) and (b)
to describe the cross sections as functions of the renormalization
and the factorization scales varied independently and
simultaneously. We show the cross section profile both at the LO and
at the QCD  NLO by adopting the event selection scheme (II) and
taking the LHT parameters $f=1~{\rm TeV}$ and $\kappa = 1$. The
curves of the $\sigma_{LO}$ and $\sigma_{NLO}$ for the $pp \to
W_H^+Z_H+X$ process are labeled by ''$LO_{+}$'' and ''$NLO_{+}$'',
while those for the $pp \to W_H^-Z_H+X$ process are labeled by
''$LO_{-}$'' and ''$NLO_{-}$'', respectively. The two figures trace
the scale dependence following a contour in the $\mu_R-\mu_F$ plane
as shown in each left panel of Figs.\ref{fig6}(a) and (b). From
these figures we can see that the QCD NLO corrections do not
obviously improve the scale uncertainty with individual variation of
either $\mu_R$ or $\mu_F$. Particularly, the LO partonic processes
for the $pp \to W_H^{\pm}Z_H+X$ processes are pure electroweak
channels where the $\mu_R$ dependence is invisible at the LO, as
shown in Figs.\ref{fig6}(a)-(3), (a)-(5), (b)-(3) and (b)-(5).
Figs.\ref{fig6}(a)-(1) and (b)-(1) show that the scale uncertainty
is reduced by the NLO corrections with simultaneous variation of
$\mu_R$ and $\mu_F$. It demonstrates that when we set $\mu_R=\mu_F$
and vary both scales simultaneously, it may lead to artificial
cancelations among renormalization and factorization logarithms, and
thus hiding the scale dependence. In the following discussions the
factorization/renormalization scale is fixed as
$\mu_0=(m_{W_H}+m_{Z_H})/2$.

\par
\subsection{Dependence on global symmetry breaking scale $f$ }
\par
\begin{figure}[htbp]
\begin{center}
\includegraphics[width=0.45\textwidth]{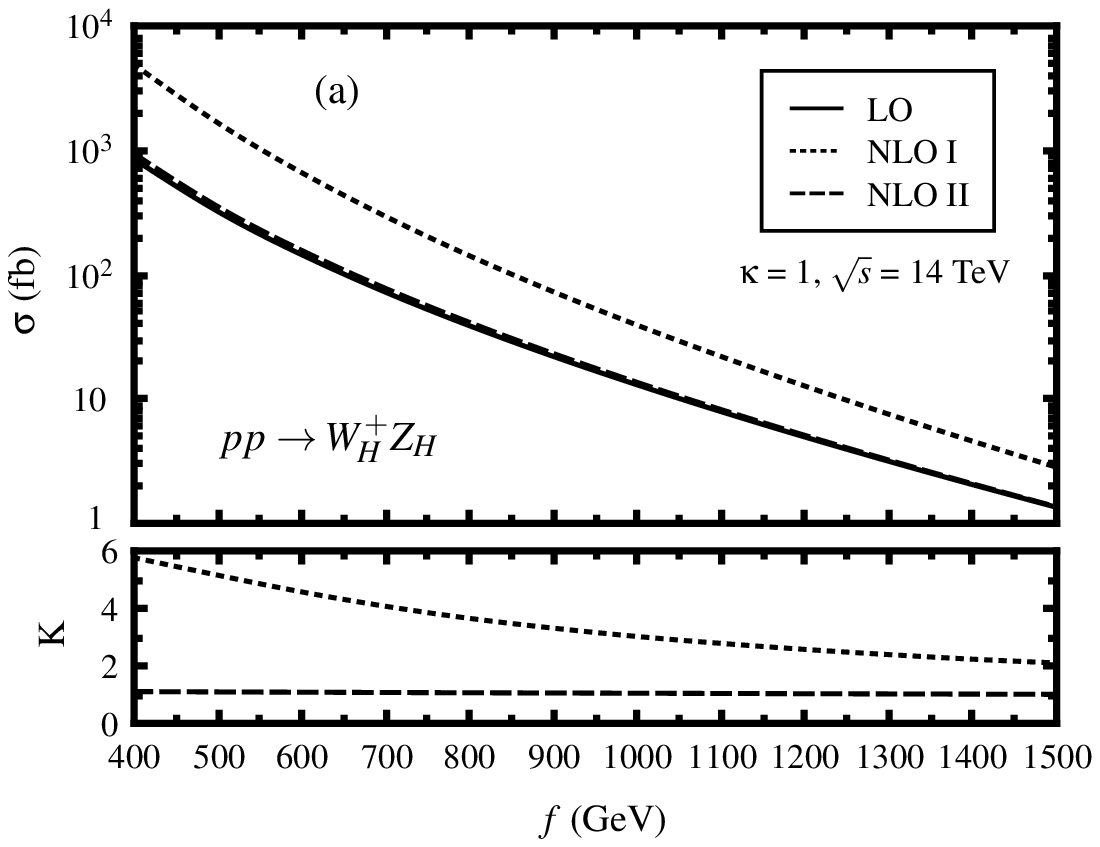}
\includegraphics[width=0.45\textwidth]{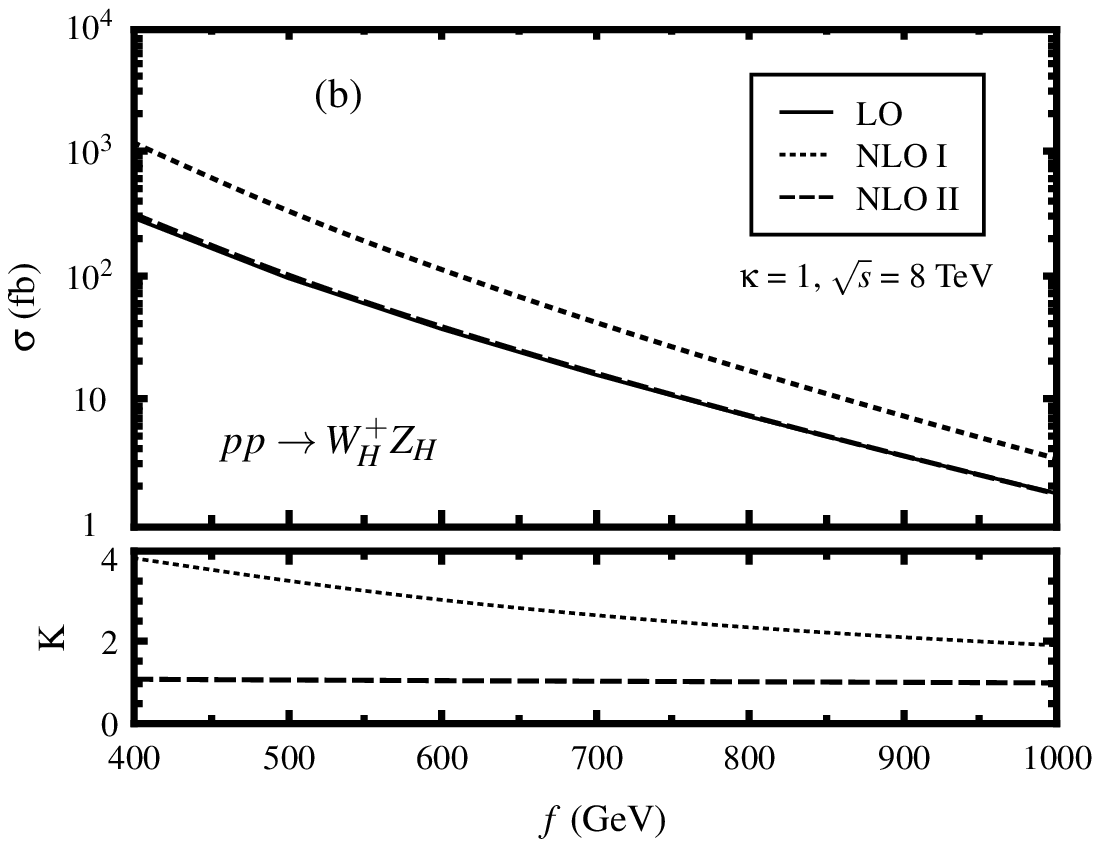}
\\~~  \\
\includegraphics[width=0.45\textwidth]{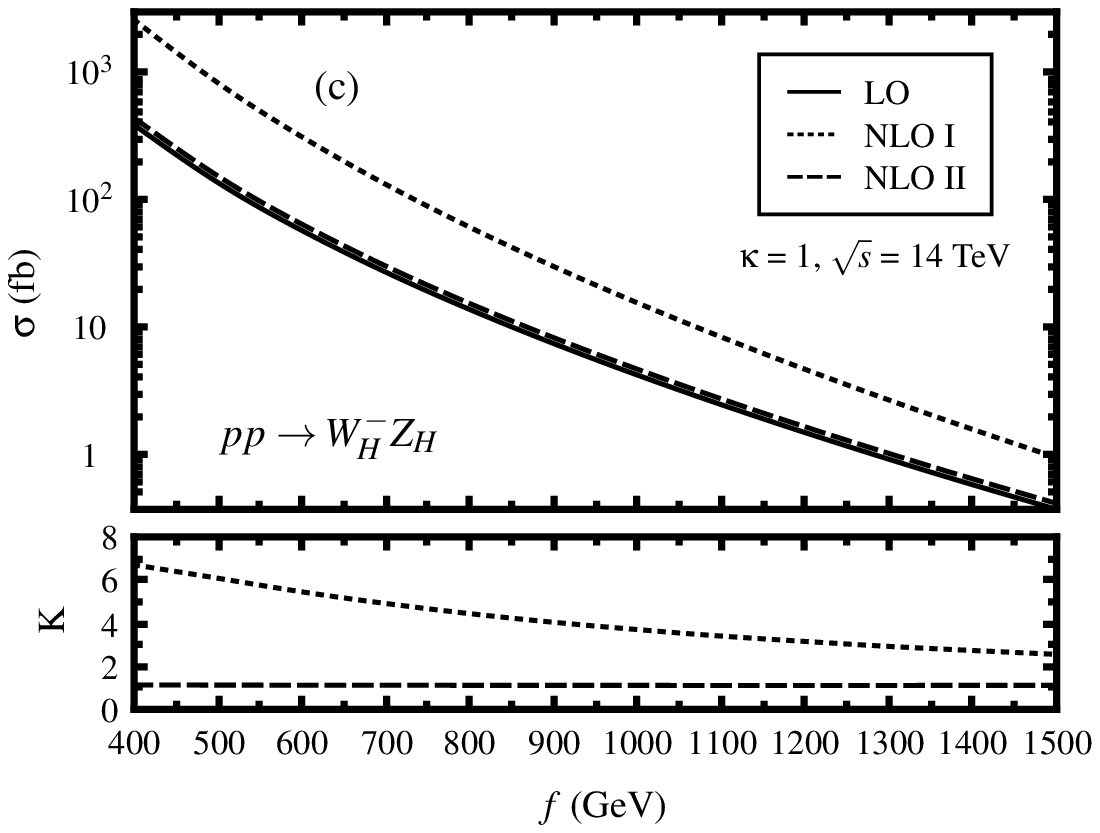}
\includegraphics[width=0.45\textwidth]{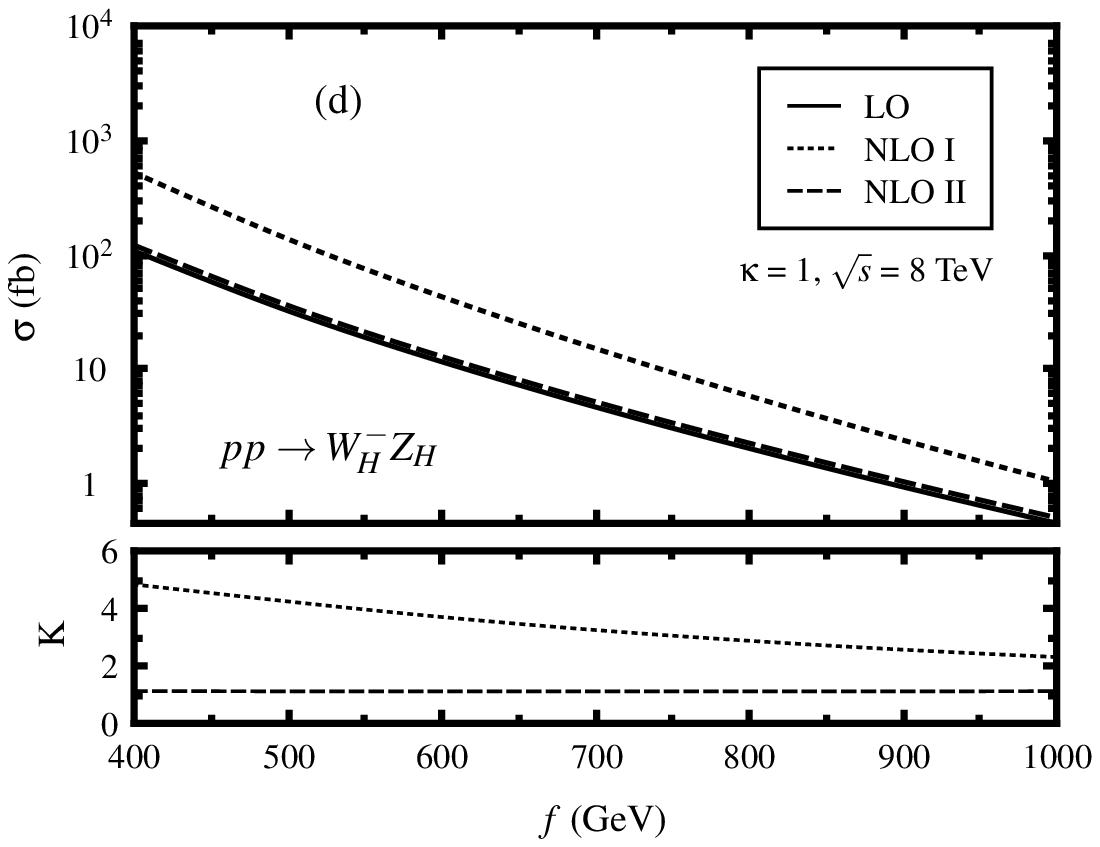}
\hspace{0in}%
\caption{\label{fig7} The LO, QCD NLO corrected integrated cross
sections and the corresponding $K$-factors as the functions of the
global symmetry breaking scale $f$ with $\kappa =1$. (a) for the $pp
\to W_H^+Z_H+X$ process at $\sqrt{s}=14~{\rm TeV}$ LHC. (b) for the
$pp \to W_H^+Z_H+X$ process at $\sqrt{s}=8~{\rm TeV}$ LHC. (c) for
the $pp \to W_H^-Z_H+X$ process at $\sqrt{s}=14~{\rm TeV}$ LHC. (d)
for the $pp \to W_H^-Z_H+X$ process at $\sqrt{s}=8~{\rm TeV}$ LHC.}
\end{center}
\end{figure}

We depict the LO, QCD NLO corrected integrated cross sections and
the corresponding $K$-factors for the $pp \to W_H^+Z_H+X$ and $pp
\to W_H^-Z_H+X$ processes as functions of the global symmetry
breaking scale $f$ at the $\sqrt{s}=14~{\rm TeV}$ and the
$\sqrt{s}=8~{\rm TeV}$ LHC in Figs.\ref{fig7}(a), (b), (c) and (d),
respectively, with $\kappa =1$. The curves labeled by "NLO I" and
"NLO II" are for the QCD NLO corrected cross sections using the (I)
and (II) selection schemes, respectively. Figs.\ref{fig7}(a,b,c,d)
demonstrate that the LO and QCD NLO corrected total cross sections
for the $pp \to W_H^{\pm}Z_H+X$ processes decrease sensitively with
the increment of $f$ due to the fact that the masses of final $W_H$
and $Z_H$ become heavier and consequently the phase space becomes
smaller as the increment of $f$. The numerical results for the $pp
\to W_H^{\pm}Z_H+X$ processes at the LHC for some typical values of
$f$ are presented in Table \ref{tab3}.
\begin{table}
\begin{center}
\begin{tabular}{c|c|c|c|c|c|c|c}
\hline
$\sqrt{s}$ & $f$  &$\sigma_{LO}^{(W^+)}$  & $\sigma_{NLO}^{(W^+)}$ &$K^{(W^+)}$ &$\sigma_{LO}^{(W^-)}$
&$\sigma_{NLO}^{(W^-)}$&$K^{(W^-)}$\\
$({\rm TeV})$ & $({\rm GeV})$ &$({\rm fb})$&$({\rm fb})$& &$({\rm fb})$& &$({\rm fb})$\\
\hline
     & 500  & 321.096(8)   & 350.8(1)     & 1.09  & 136.130(5)   & 153.12(7)  & 1.12      \\
     & 700  & 72.055(2)    & 77.26(3)     & 1.07  & 26.888(1)    & 30.03(1)   & 1.12      \\
 14  & 900  & 21.9589(5)   & 23.159(7)    & 1.05  & 7.3867(3)    & 8.216(3)   & 1.11      \\
     & 1100 & 7.8997(2)    & 8.203(3)     & 1.04  & 2.44038(9)   & 2.709(1)   & 1.11      \\
\hline
    & 500   & 94.168(2)    & 100.01(5)    & 1.06  & 32.379(1)    & 36.10(4)   & 1.11      \\
  8 & 700   & 15.7549(4)   & 16.259(7)    & 1.03  & 4.6757(2)    & 5.192(6)   & 1.11      \\
    & 900   & 3.49785(8)   & 3.515(1)     & 1.01  & 0.93703(3)   & 1.044(1)   & 1.11      \\
\hline
\end{tabular}
\end{center}
\begin{center}
\begin{minipage}{15cm}
\caption{\label{tab3} The numerical results of $\sigma_{LO}^{W^+}$,
$\sigma_{NLO}^{W^+}$, $\sigma_{NLO}^{W^-}$, $\sigma_{NLO}^{W^-}$ for
the $pp \to W_H^{+}Z_H+X$ and $pp \to W_H^{-}Z_H+X$ processes and
their corresponding $K$-factors at the $\sqrt{s}=14~{\rm TeV}$ and
the $\sqrt{s}=8~{\rm TeV}$ LHC by adopting the event selection
scheme (II) and taking $\kappa = 1$, $\mu=\mu_0$ for some typical
values of $f$.  }
\end{minipage}
\end{center}
\end{table}

\par
\subsection{Differential cross sections }
\par
In this subsection we focus on the kinematic distributions of final
decay products. The $W_H Z_H$ associated production at the LHC are
followed by the heavy gauge boson decays of $W_H^{\mp} \to W^{\mp}
A_H \to \mu^{\mp} \stackrel{(-)}{\nu_{\mu}} A_H$ and $Z_H \to H
A_H$. The branching ratios of decays for the $W_H$ boson, $Z_H$
boson and $W$ boson are taken as $Br(W_H \to W A_H)=100\%$, $Br(Z_H
\to H A_H)=100\%$ for $\kappa=1$ and $f=1~{\rm TeV}$
\cite{cpyuan:2006ph} and $Br(W^{\mp} \to \mu^{\mp}
\stackrel{(-)}{\nu_{\mu}})=10.57\%$ \cite{databook}, respectively.
In the following we consider the $W_H Z_H$ production channel
including its subsequential decays as
\begin{eqnarray}\label{channel}
pp \to W_H^{\mp} Z_H \to W^{\mp} A_H H A_H \to \mu^{\mp}
\stackrel{(-)}{\nu_{\mu}} A_H H A_H.
\end{eqnarray}
Thus one expects that the $W_H Z_H$ production at the LHC could be
detected via the $\mu^{\mp} H + \slashed{E}_T$ ($\slashed{E}_T$ =
transverse energy of $\stackrel{(-)}{\nu_{\mu}} + 2 A_H$) channel.

\par
\begin{figure}[htbp]
\begin{center}
\includegraphics[width=0.45\textwidth]{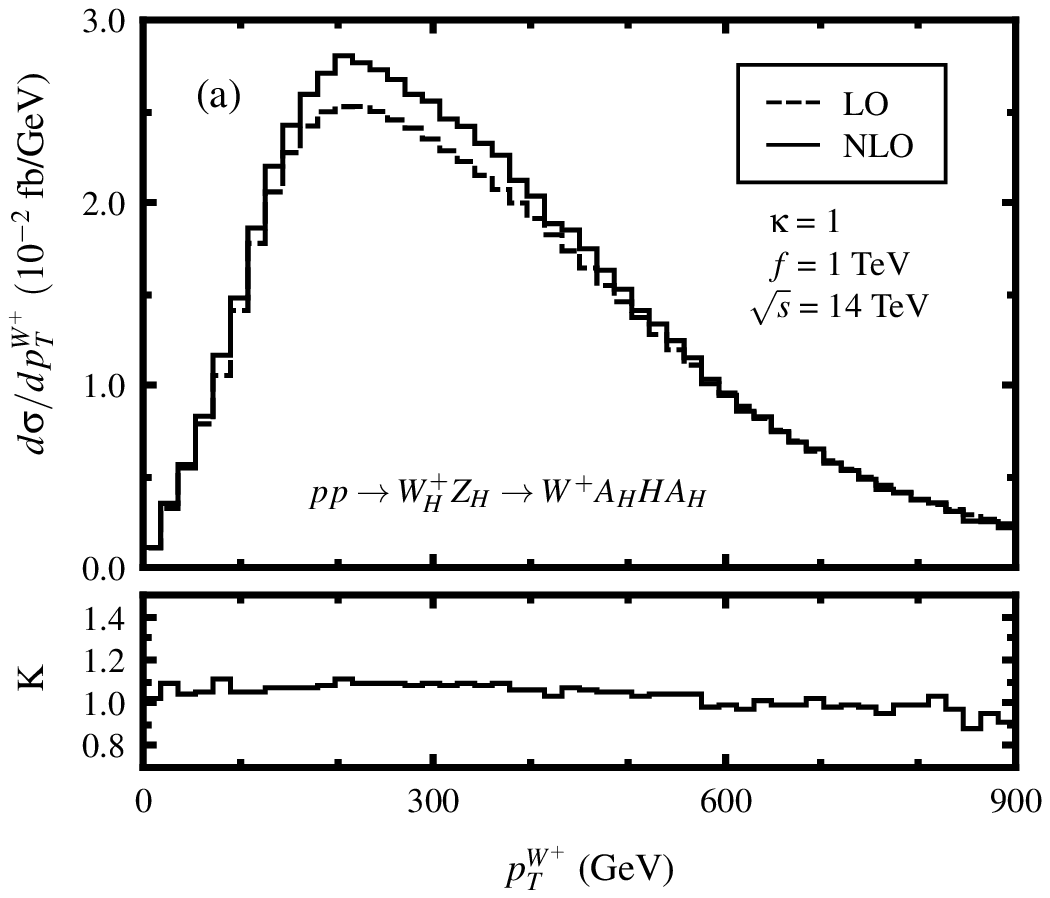}
\includegraphics[width=0.45\textwidth]{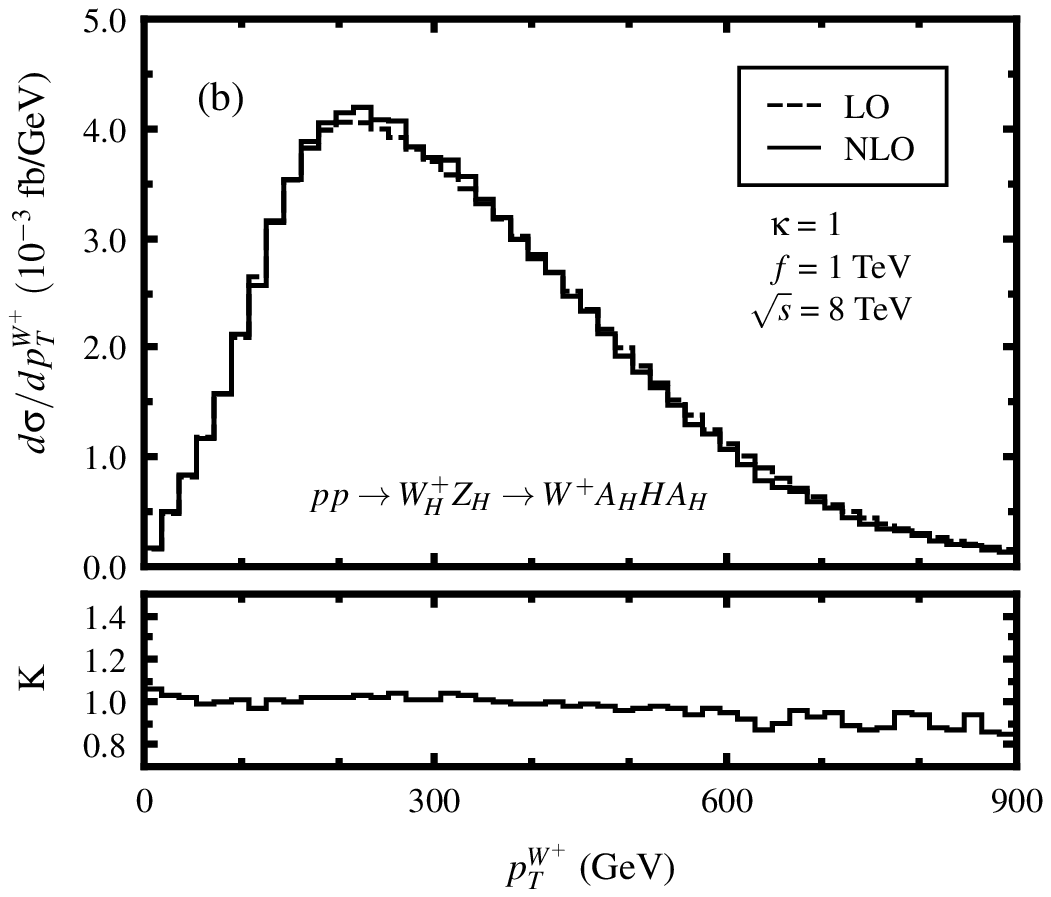}
\\~~  \\
\includegraphics[width=0.45\textwidth]{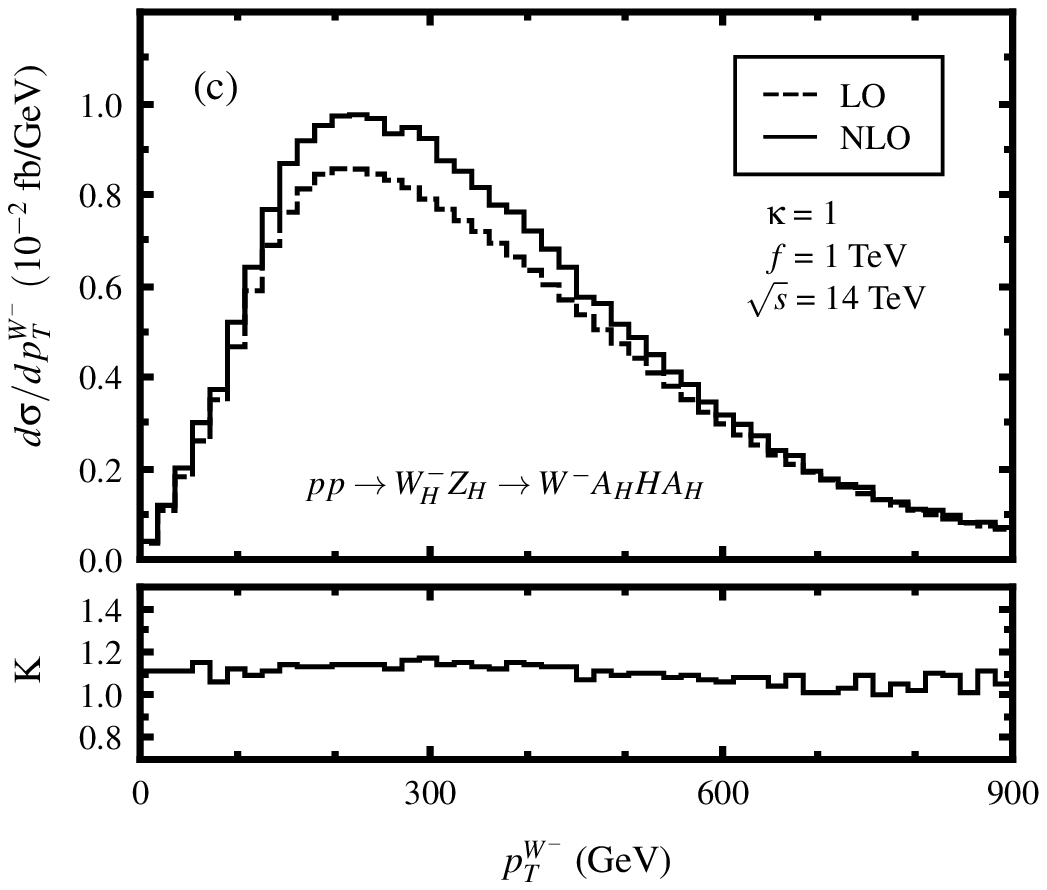}
\includegraphics[width=0.45\textwidth]{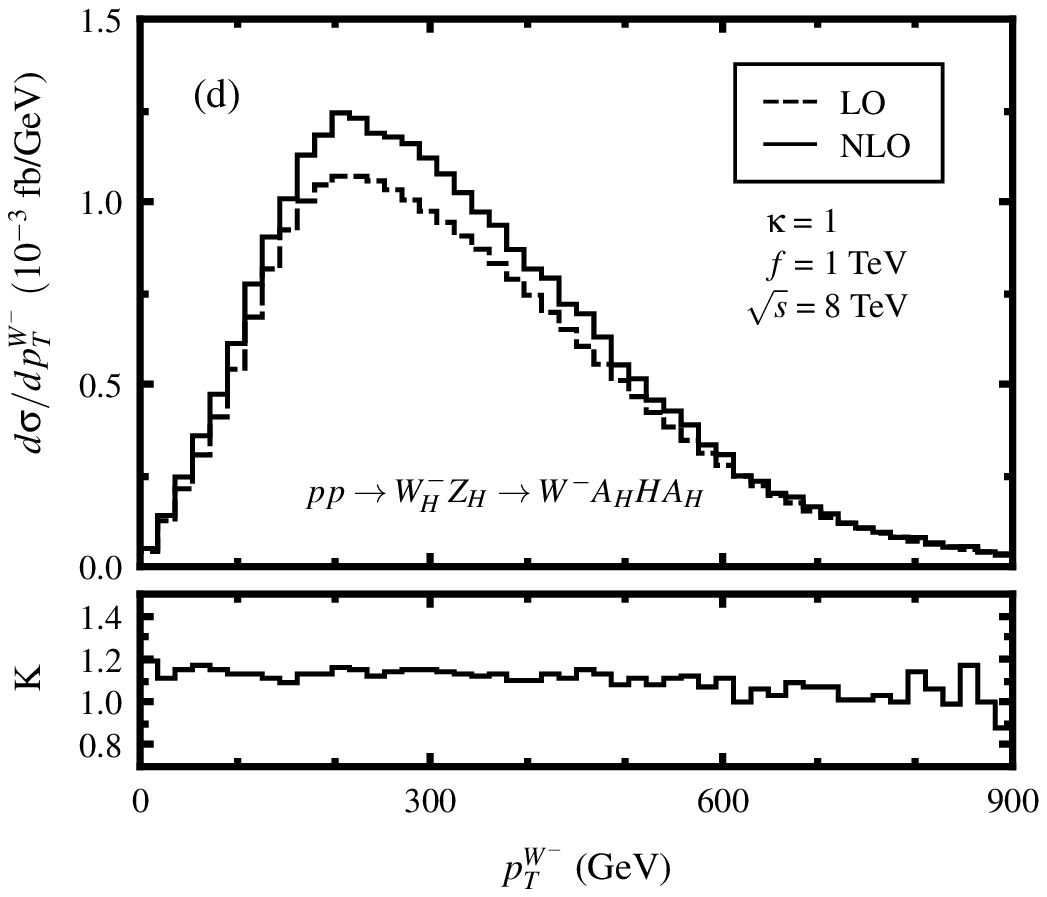}
\caption{\label{fig8} The LO, QCD NLO corrected $p_T^{W}$
distributions and the corresponding $K$-factors in scheme (II) for
the \ppfinala~processes by taking $f=1~{\rm TeV}$ and $\kappa=1$.
(a) for the $pp \to W^+_H Z_H \to W^+ A_H H A_H + X$ process at the
$\sqrt{s}=14~{\rm TeV}$ LHC. (b) for the $pp \to W^+_H Z_H \to W^+
A_H H A_H + X$ process at the $\sqrt{s}=8~{\rm TeV}$ LHC. (c) for
the $pp \to W^-_H Z_H \to W^- A_H H A_H + X$ process at the
$\sqrt{s}=14~{\rm TeV}$ LHC. (d) for the $pp \to W^-_H Z_H \to W^-
A_H H A_H + X$ process at the $\sqrt{s}=8~{\rm TeV}$ LHC. }
\end{center}
\end{figure}
\begin{figure}[htbp]
\begin{center}
\includegraphics[width=0.45\textwidth]{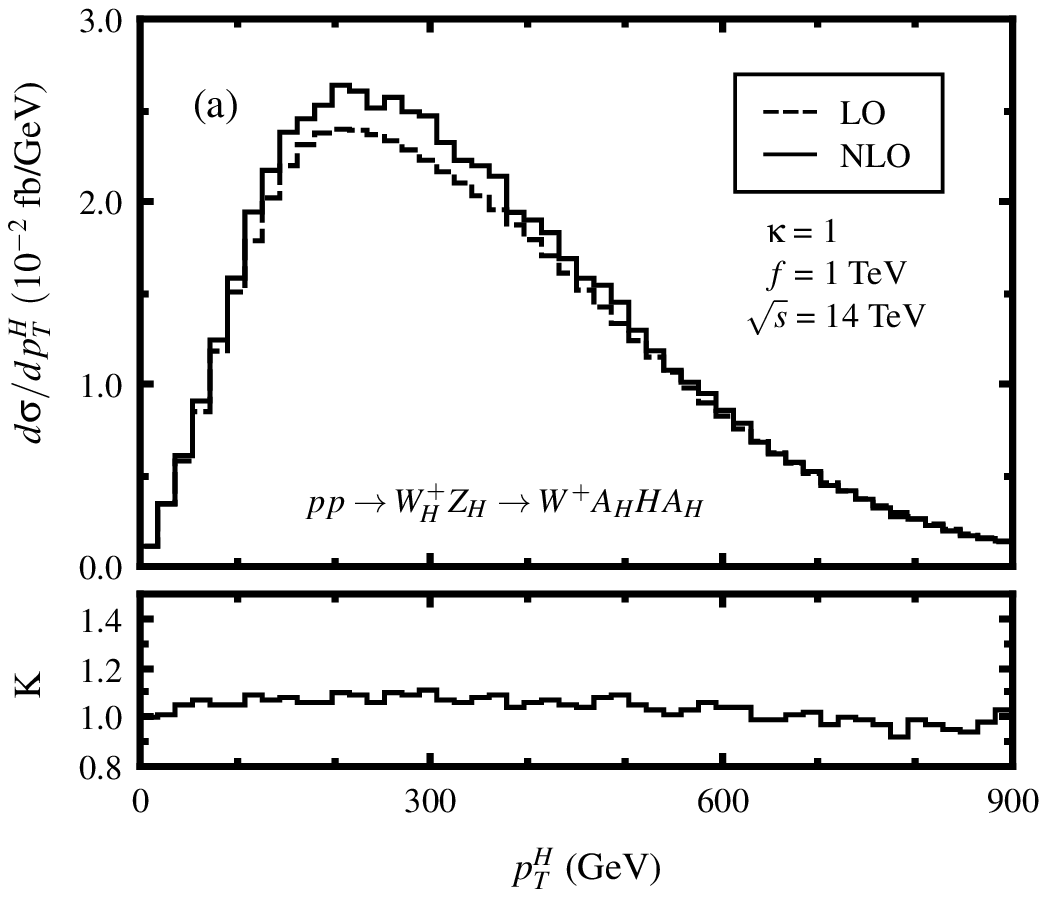}
\includegraphics[width=0.45\textwidth]{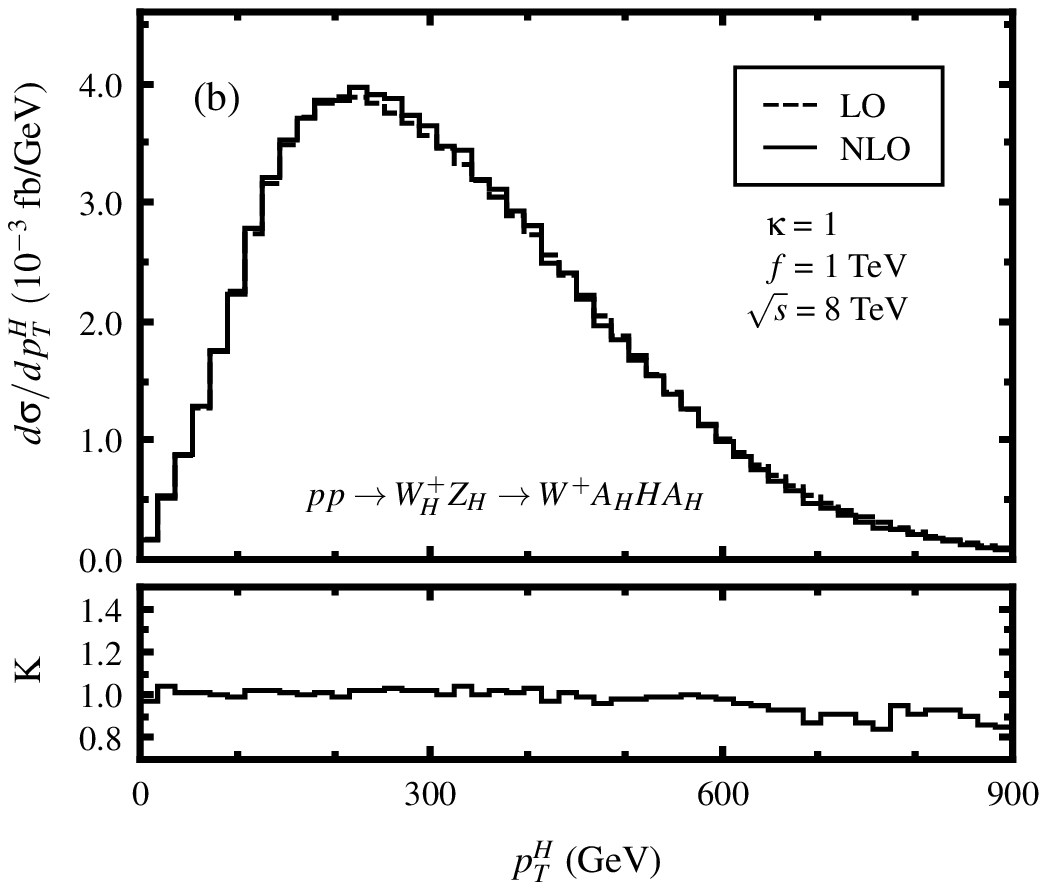}
\\~~  \\
\includegraphics[width=0.45\textwidth]{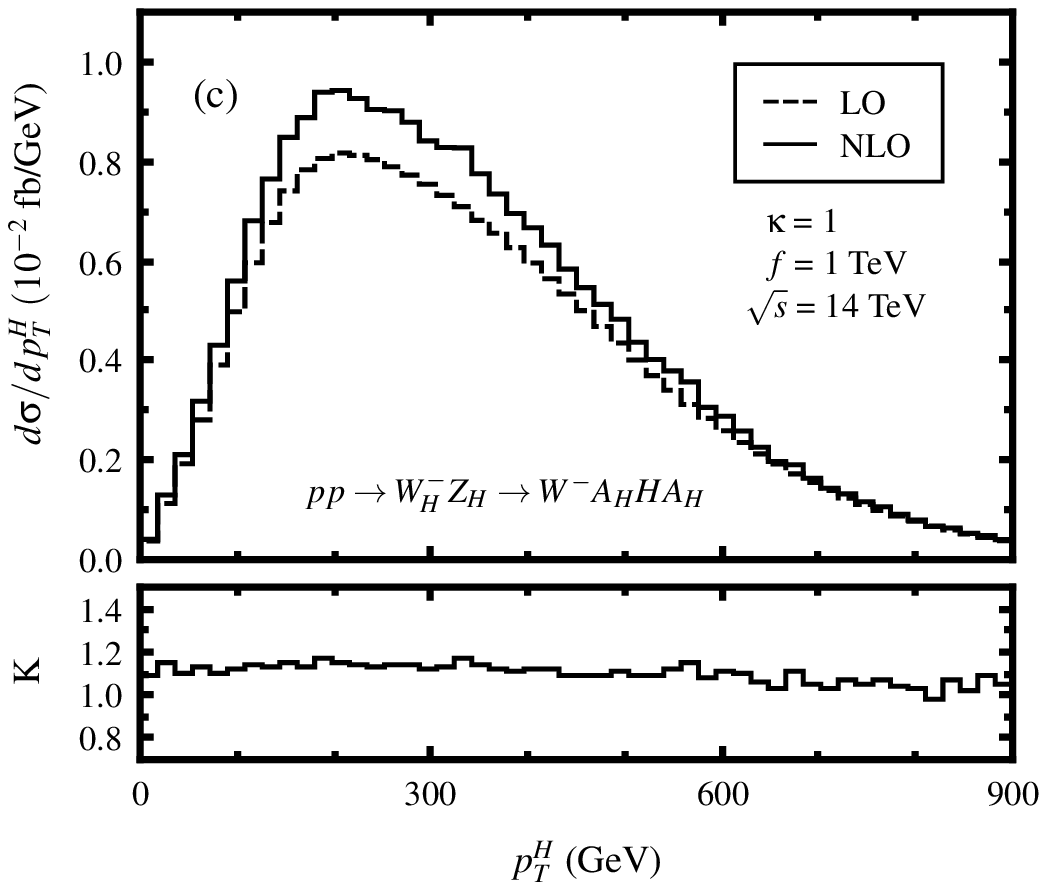}
\includegraphics[width=0.45\textwidth]{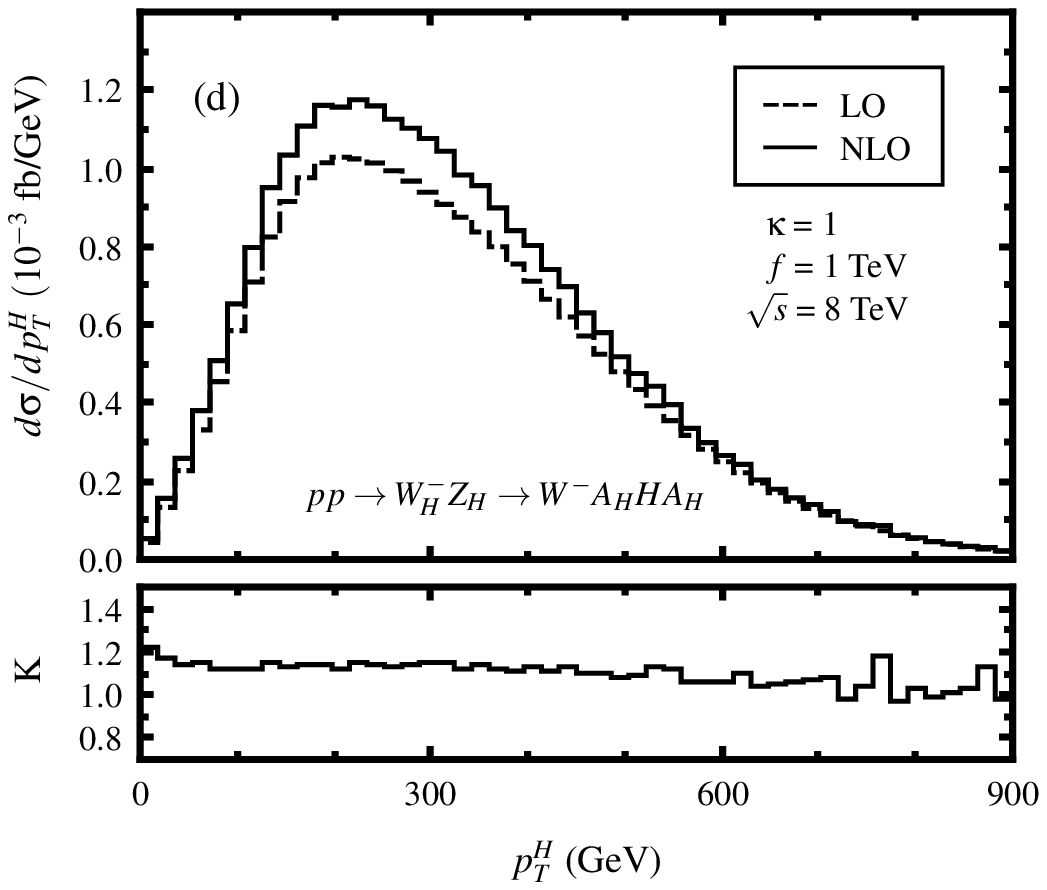}
\caption{\label{fig9} The LO, QCD NLO corrected $p_T^{H}$
distributions and the corresponding $K$-factors in scheme (II) for
the \ppfinala~processes by taking $f=1~{\rm TeV}$ and $\kappa=1$.
(a) for the \ppfinalap~ process at $\sqrt{s}=14~{\rm TeV}$ LHC. (b)
for the \ppfinalap~ process at $\sqrt{s}=8~{\rm TeV}$ LHC. (c) for
the \ppfinalam~ process at $\sqrt{s}=14~{\rm TeV}$ LHC. (d) for the
\ppfinalam~ process at $\sqrt{s}=8~{\rm TeV}$ LHC. }
\end{center}
\end{figure}

The LO, QCD NLO corrected transverse momentum distributions of $W$
boson and the light neutral Higgs boson $H$ for the $pp \to
W_H^+Z_H+X$ and $pp \to W_H^-Z_H+X$ processes, and the corresponding
$K$-factors in scheme (II) at the $\sqrt{s}=14~{\rm TeV}$ LHC and
the $\sqrt{s}=8~{\rm TeV}$ LHC are presented in
Figs.\ref{fig8}(a,b,c,d) and Figs.\ref{fig9}(a,b,c,d) separately.
There we take $f=1~{\rm TeV}$ and $\kappa=1$. From these four
figures we find that the QCD NLO corrections enhance the LO
transverse momentum distributions in most plotted ranges of $p_{T}$,
and the $K$-factors are all less than $1.20$. The maxima of the
distributions $\frac{d \sigma_{LO,NLO}}{d p_{T}^W}$ and $\frac{d
\sigma_{LO,NLO}}{d p_{T}^H}$ are all located at about $p_{T} \sim
220~{\rm GeV}$.

\par
\begin{figure}[htbp]
\begin{center}
\includegraphics[width=0.45\textwidth]{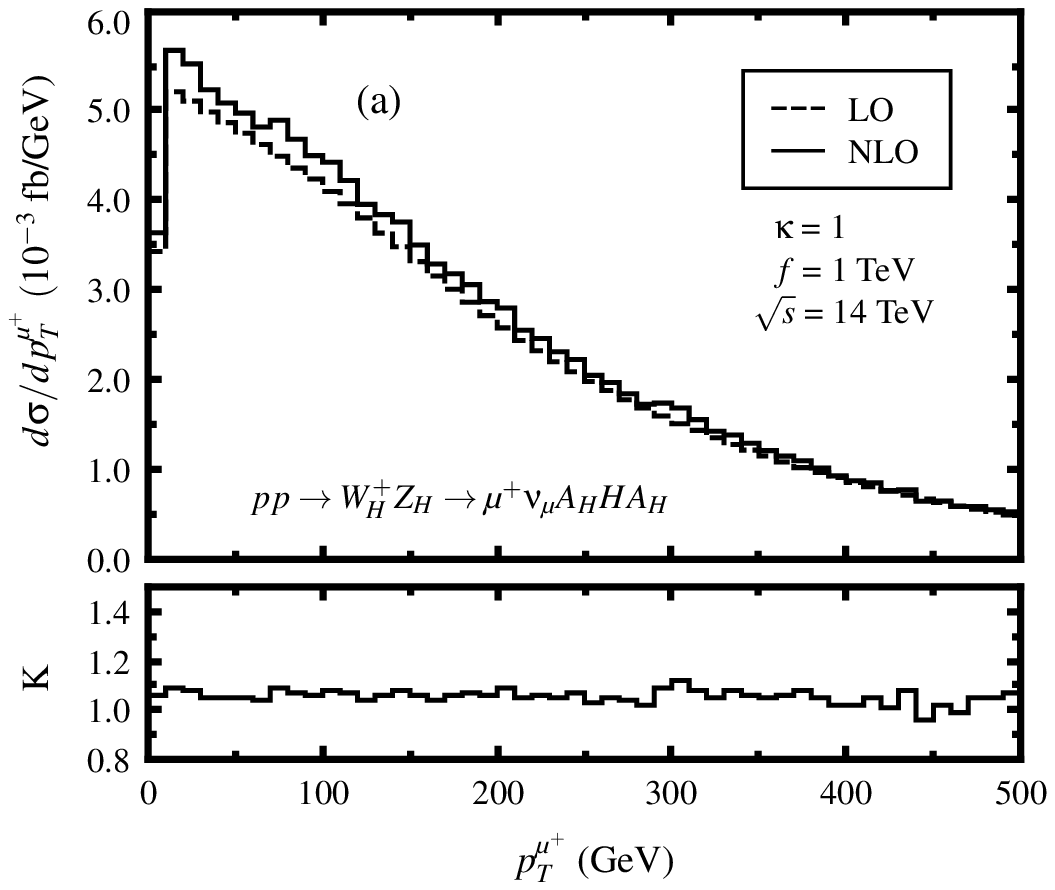}
\includegraphics[width=0.45\textwidth]{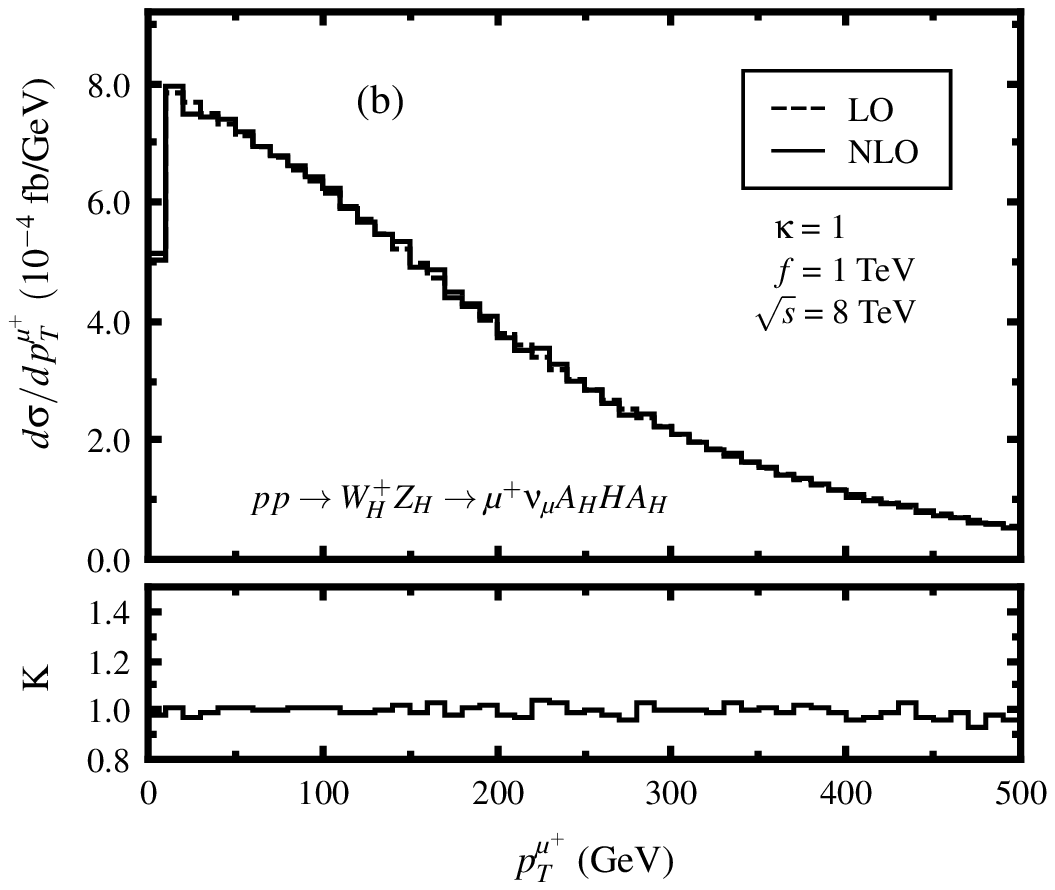}
\\~~  \\
\includegraphics[width=0.45\textwidth]{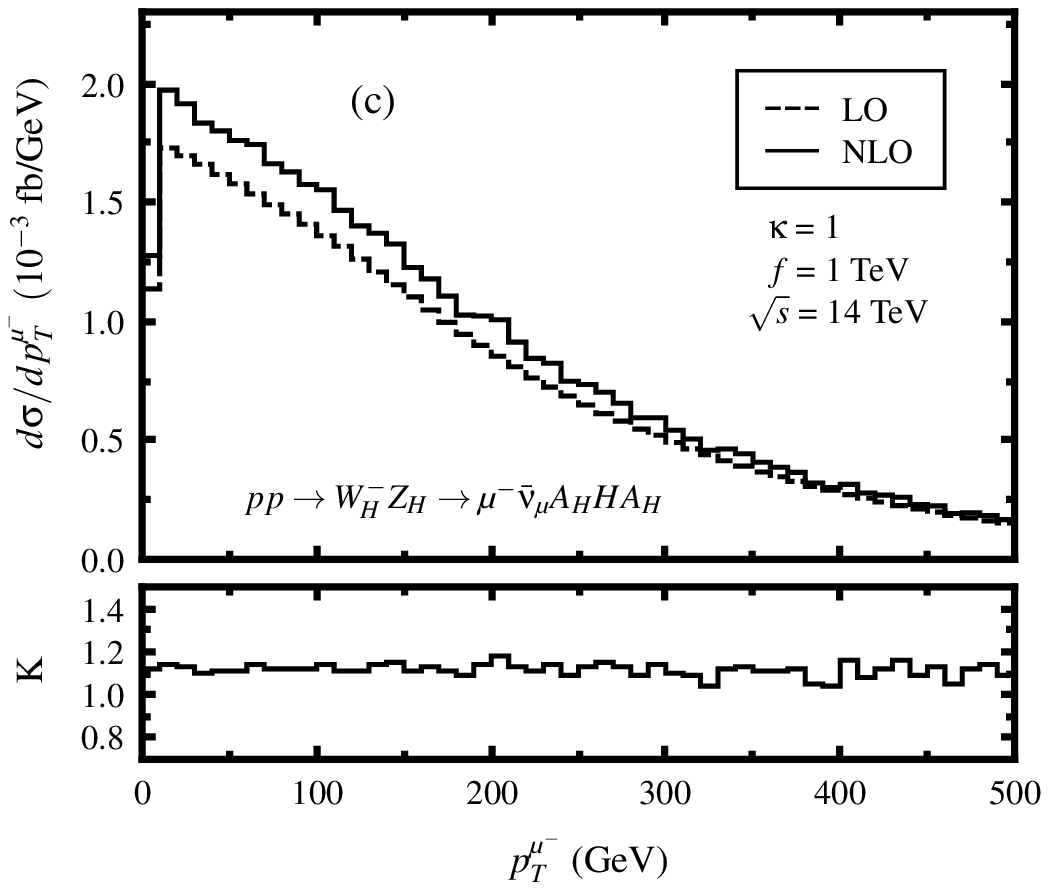}
\includegraphics[width=0.45\textwidth]{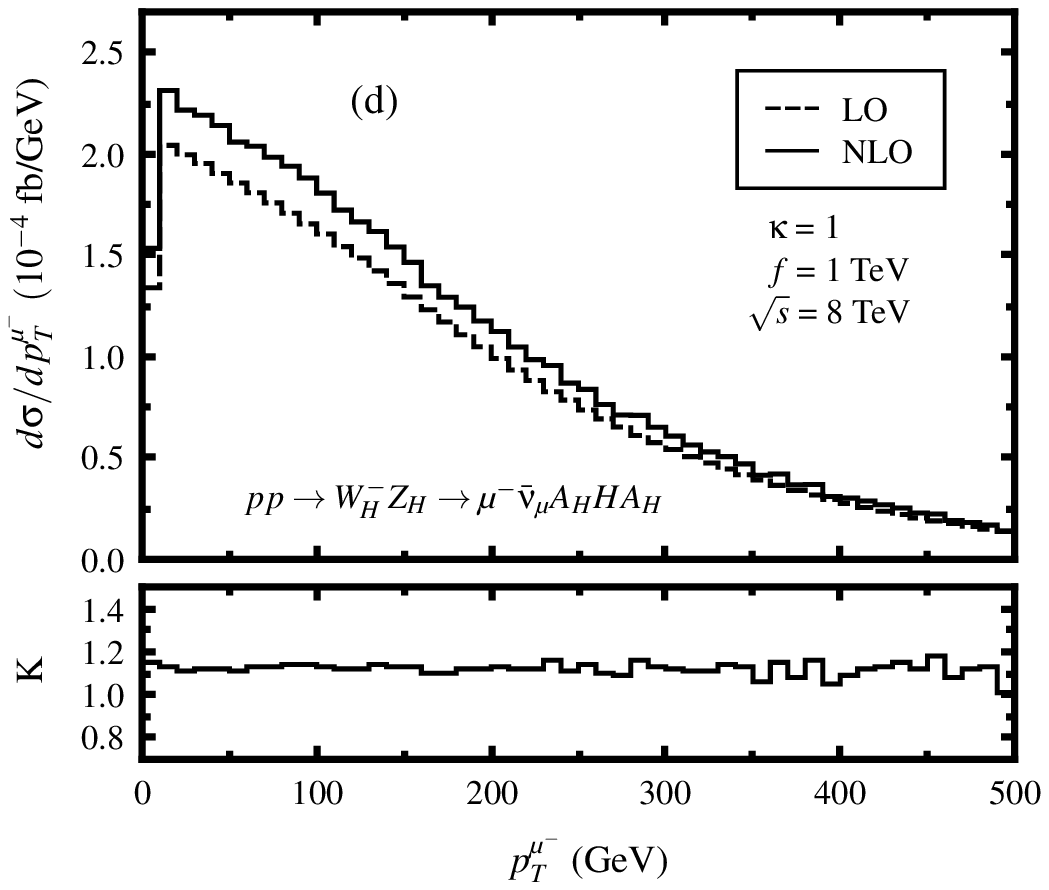}
\caption{\label{fig10} The LO, QCD NLO corrected $p_T^{\mu}$
distributions and the corresponding $K$-factors in scheme (II) for
the $pp \to W_H^{\pm}Z_H \to \mu^{\pm} \stackrel{(-)}{\nu_{\mu}} A_H
H A_H+X$ processes by taking $f=1~{\rm TeV}$ and $\kappa=1$. (a) for
the \ppfinalbp~ process at $\sqrt{s}=14~{\rm TeV}$ LHC. (b) for the
\ppfinalbp~ process at $\sqrt{s}=8~{\rm TeV}$ LHC. (c) for the
\ppfinalbm~ process at $\sqrt{s}=14~{\rm TeV}$ LHC. (d) for the
\ppfinalbm~ process at $\sqrt{s}=8~{\rm TeV}$ LHC.  }
\end{center}
\end{figure}
\begin{figure}[htbp]
\begin{center}
\includegraphics[width=0.45\textwidth]{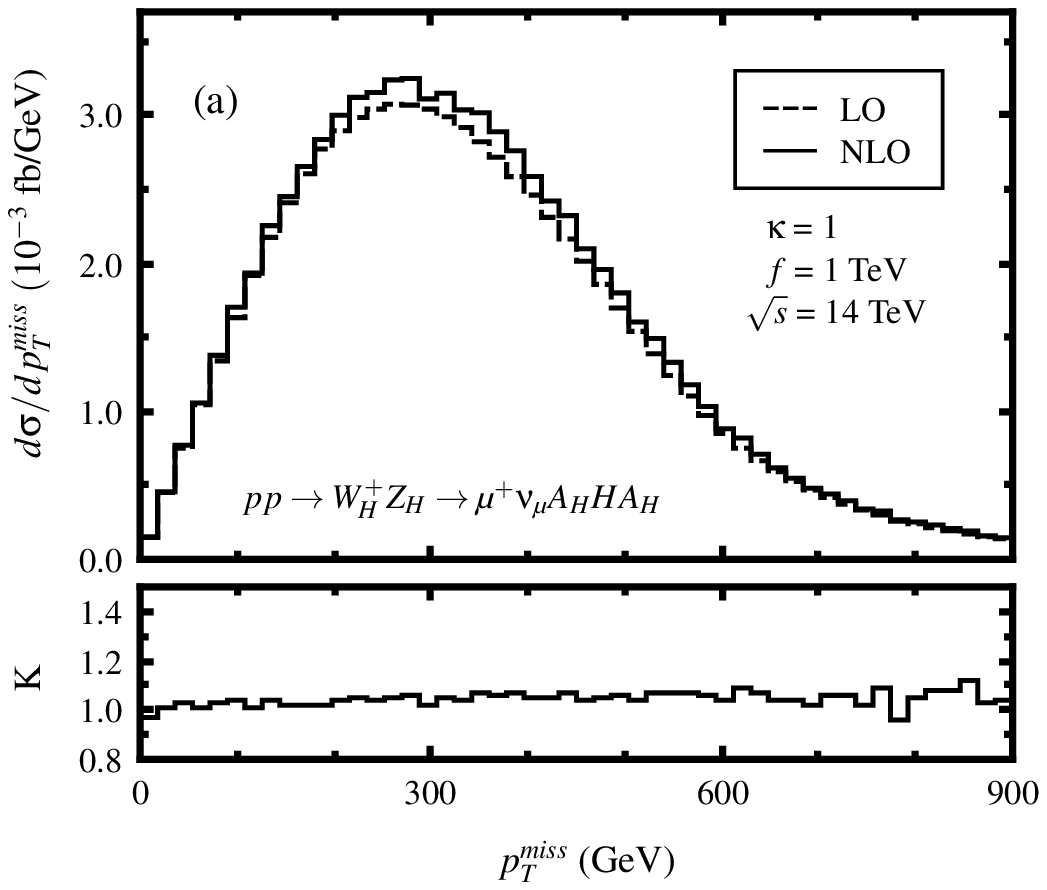}
\includegraphics[width=0.45\textwidth]{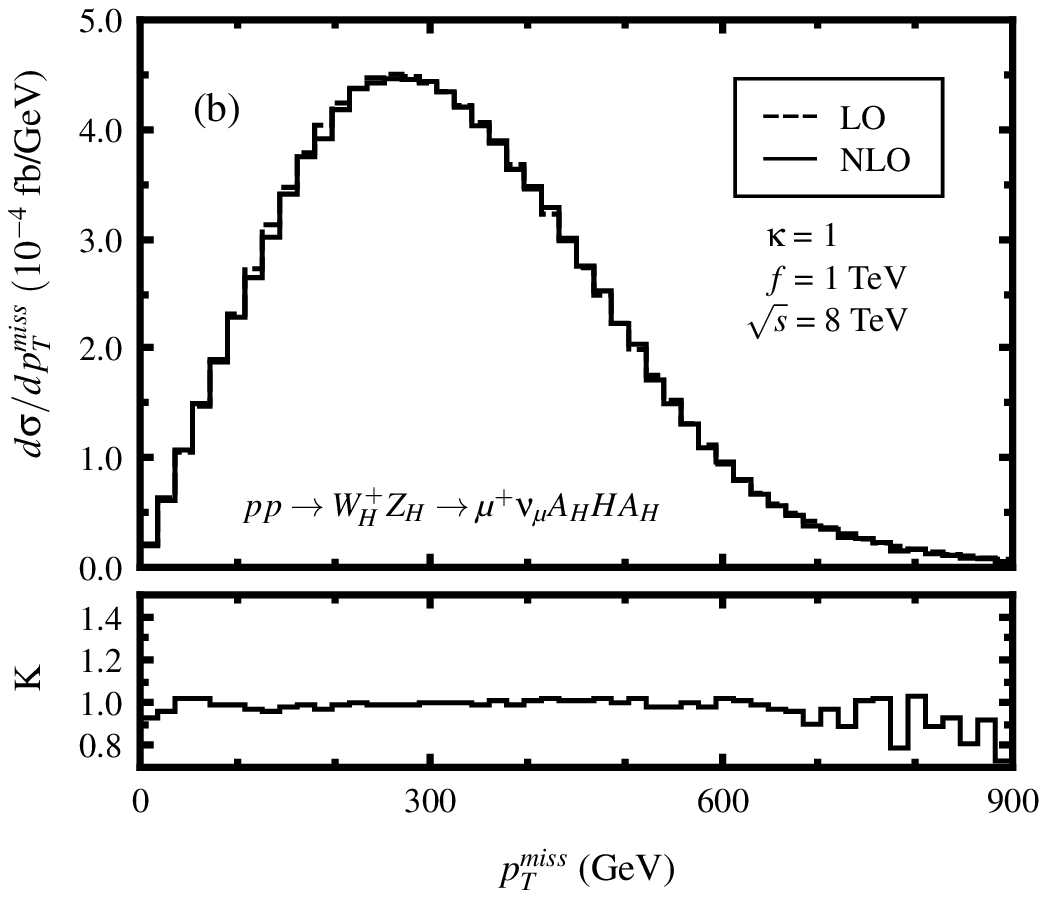}
\\~~  \\
\includegraphics[width=0.45\textwidth]{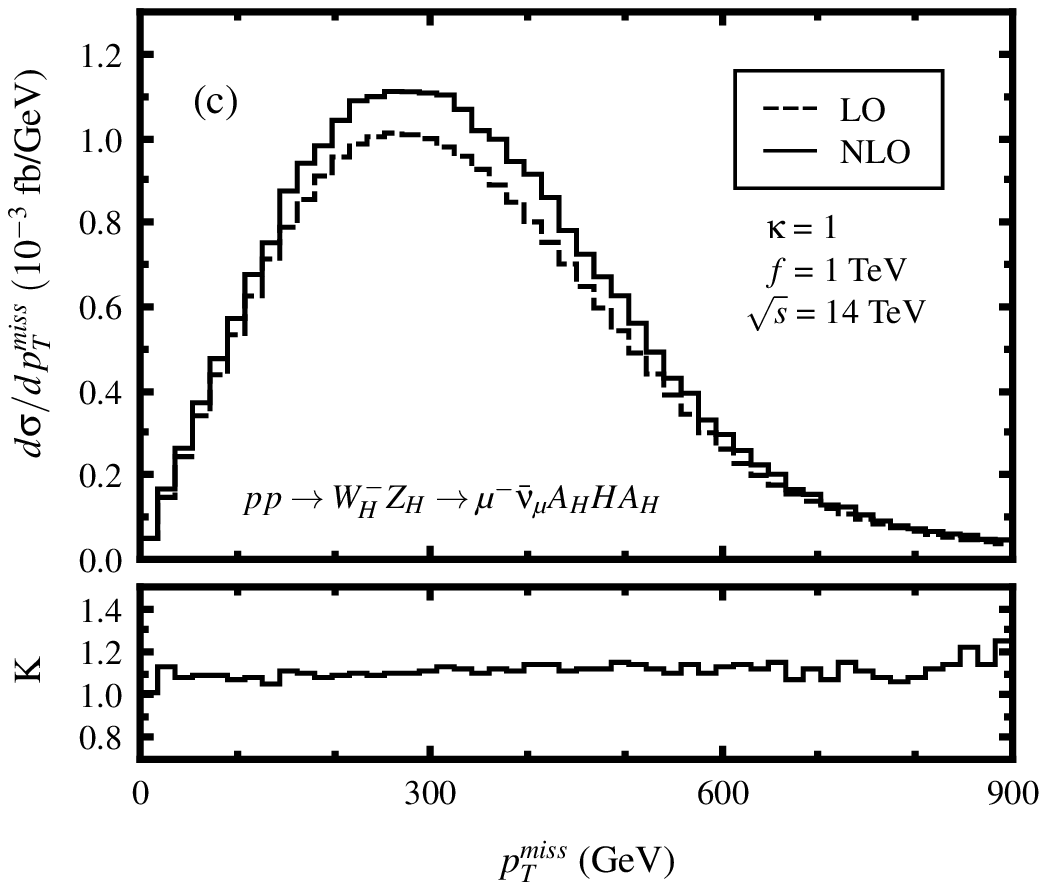}
\includegraphics[width=0.45\textwidth]{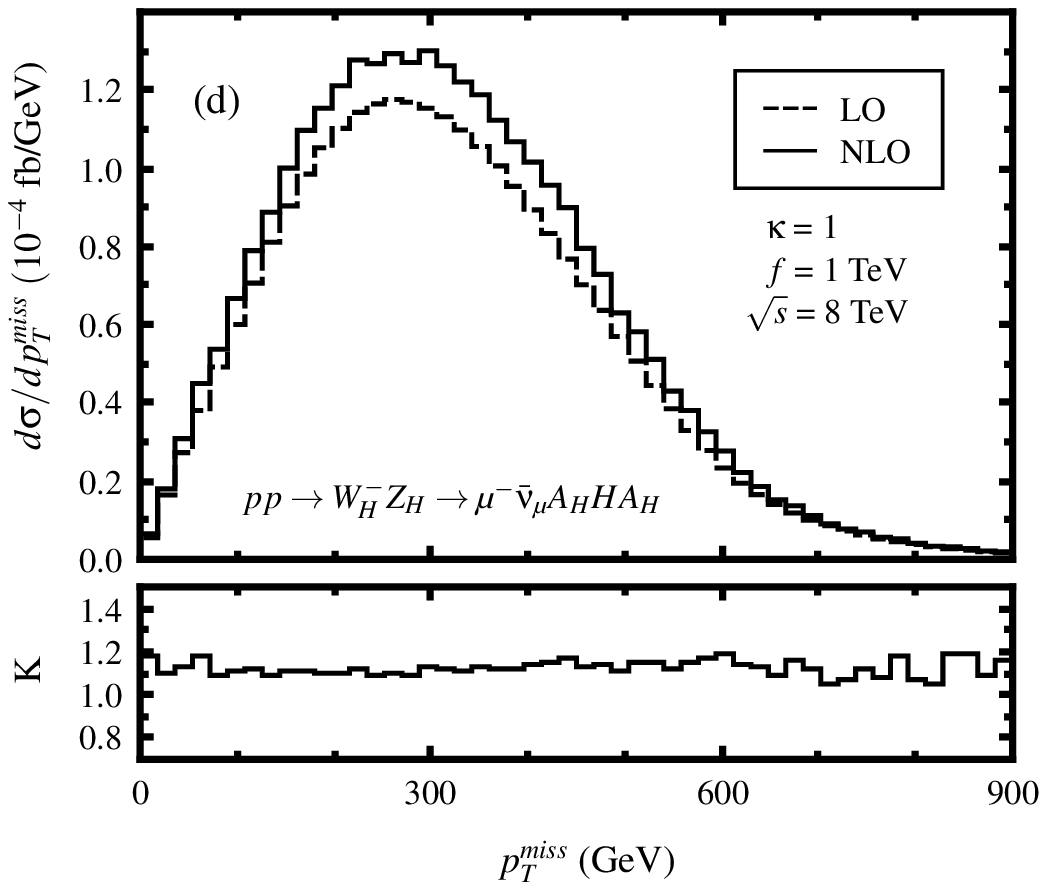}
\caption{\label{fig11} The LO, QCD NLO corrected $p_T^{miss}$
distributions and the corresponding $K$-factors in scheme (II) for
the $pp \to W_H^{\pm}Z_H \to \mu^{\pm} \stackrel{(-)}{\nu_{\mu}} A_H
H A_H+X$ processes by taking $f=1~{\rm TeV}$ and $\kappa=1$. (a) for
the \ppfinalbp~ process at the $\sqrt{s}=14~{\rm TeV}$ LHC. (b) for
the \ppfinalbp~ process at the $\sqrt{s}=8~{\rm TeV}$ LHC. (c) for
the \ppfinalbm~ process at the $\sqrt{s}=14~{\rm TeV}$ LHC. (d) for
the \ppfinalbm~ process at the $\sqrt{s}=8~{\rm TeV}$ LHC. }
\end{center}
\end{figure}

The LO and QCD NLO corrected transverse momentum distributions of
the final $\mu$-lepton and missing energy
($A_HA_H\stackrel{(-)}{\nu_{\mu}}$) for the $pp \to W_H^{-}Z_H \to
\mu^{-} \bar{\nu} A_H H A_H+X$ and $pp \to W_H^{+}Z_H \to \mu^{+}
\nu_{\mu} A_H H A_H+X$ processes, and the corresponding $K$-factors
in scheme (II) at the early LHC and the future LHC are depicted in
Figs.\ref{fig10}(a,b,c,d) and Figs.\ref{fig11}(a,b,c,d),
respectively. There we take $f=1~{\rm TeV}$ and $\kappa=1$.
Figs.\ref{fig10}(a,b) are for the $p_T$ distributions of $\mu^-$,
and Figs.\ref{fig10}(c,d) for $\mu^+$, respectively.
Figs.\ref{fig10} (a), (b), (c) and (d) demonstrate that both the LO
and the QCD NLO corrected $p_T^{\mu}$ distributions at both the
early LHC and the future LHC decrease rapidly with the increment of
$p_T^{\mu}$. Figs.\ref{fig11}(a,b,c,d) show that the LO and NLO
missing transverse momentum distributions reach their maxima at
$p_T^{miss} \sim 290~{\rm GeV}$.

\par
\begin{figure}[htbp]
\begin{center}
\includegraphics[width=0.45\textwidth]{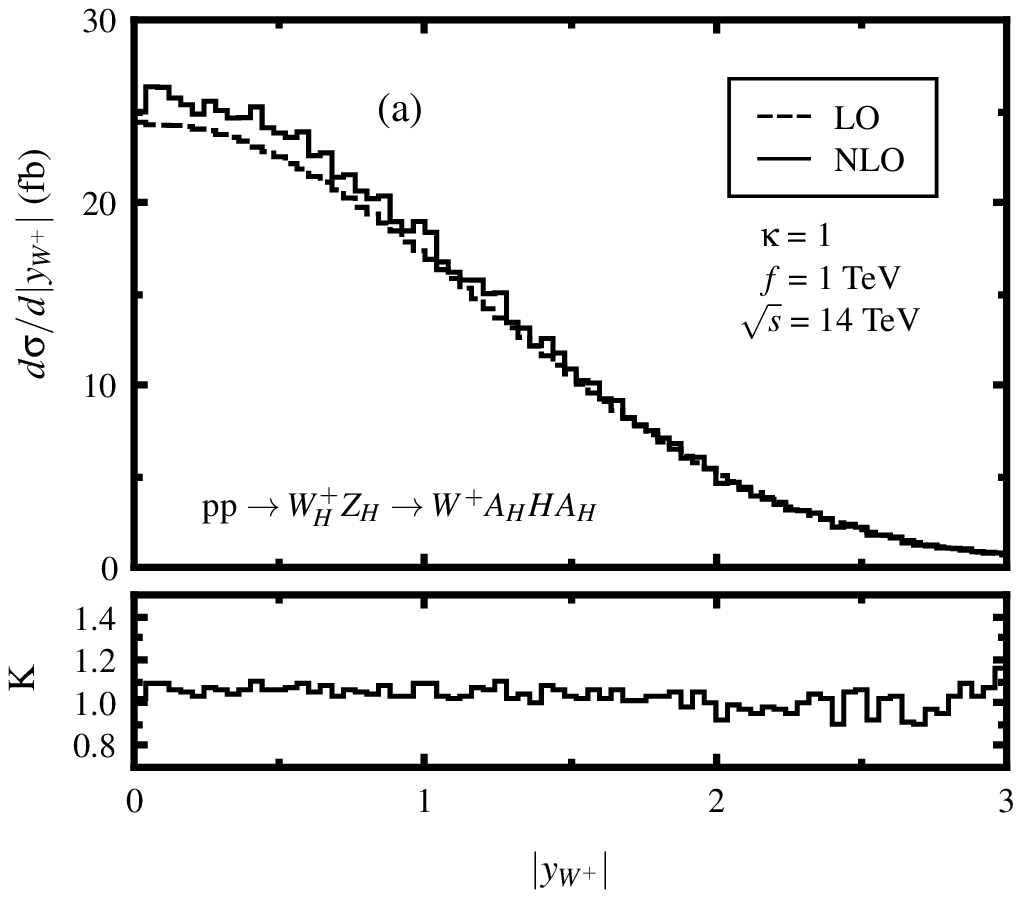}
\includegraphics[width=0.45\textwidth]{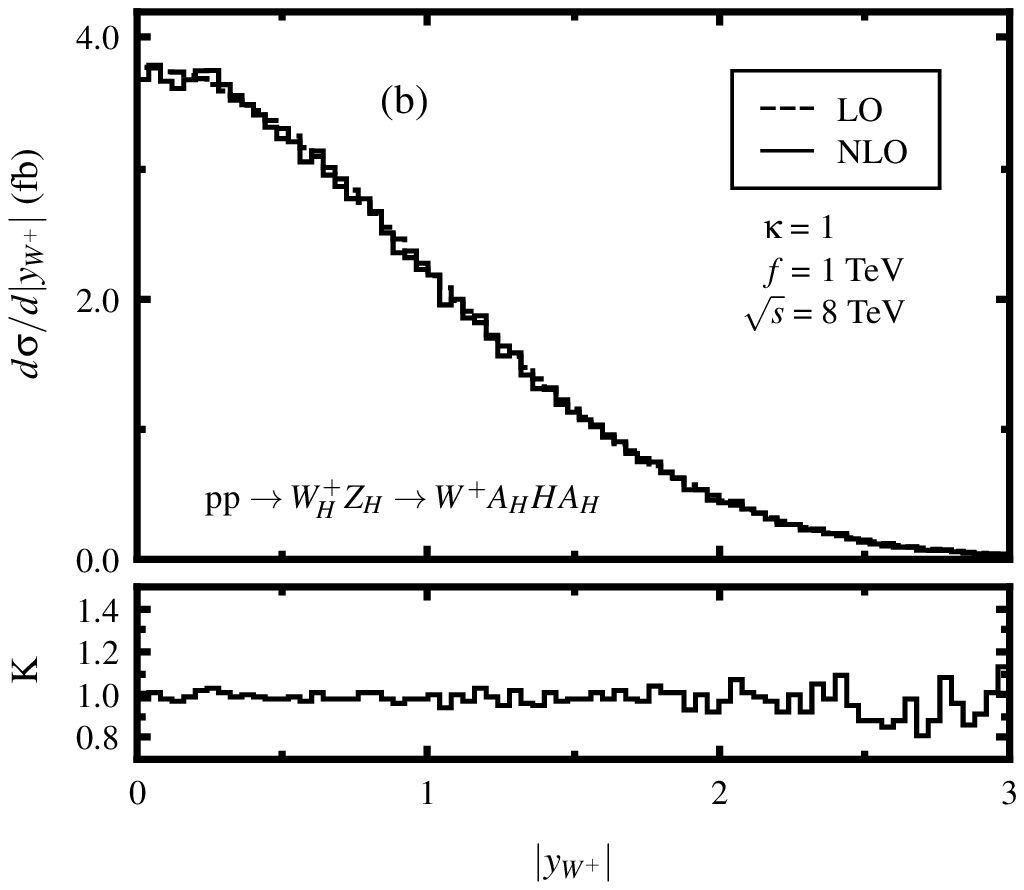}
\\~~  \\
\includegraphics[width=0.45\textwidth]{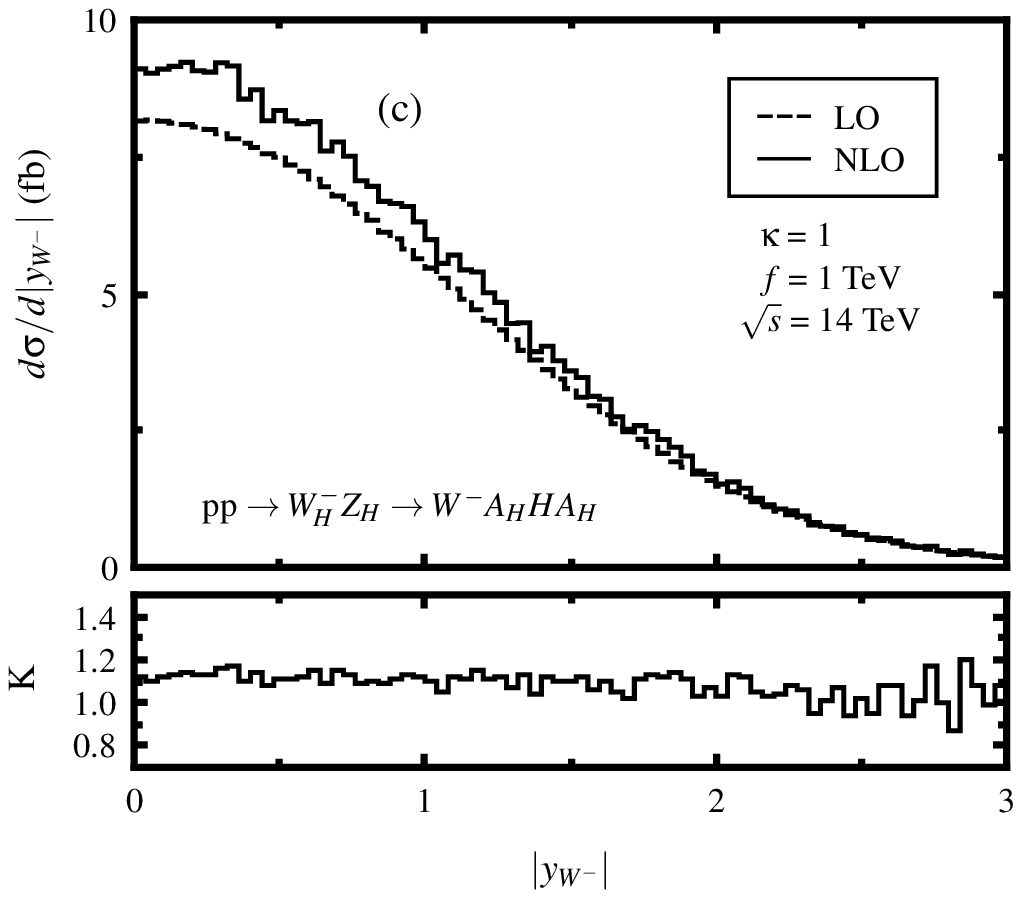}
\includegraphics[width=0.45\textwidth]{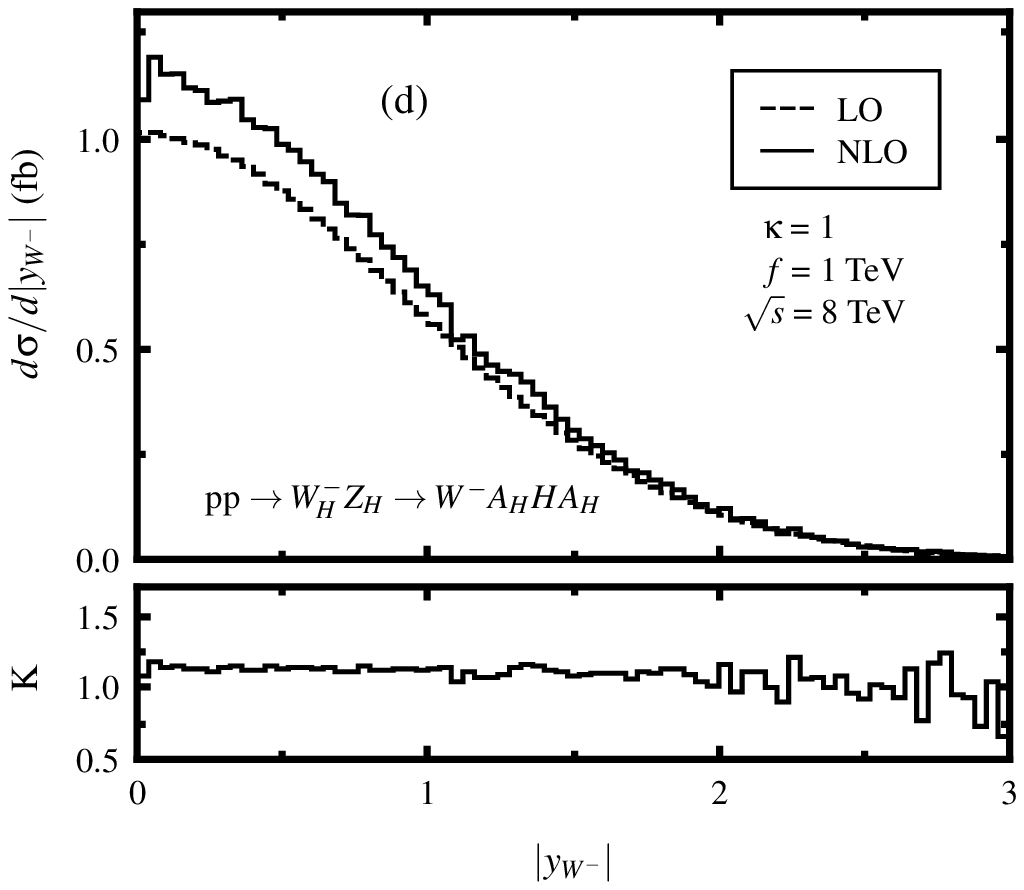}
\caption{\label{fig12} The LO, QCD NLO corrected rapidity
distributions of the of final $W$-boson, $|y_{W}|$, and the
corresponding $K$-factors in scheme (II) for the $pp \to W^{\pm}_H
Z_H \to W^{\pm} A_H H A_H + X$ processes by taking $f=1~{\rm TeV}$
and $\kappa=1$. (a) for the $pp \to W_H^{+} Z_H \to W^+A_H H A_H+X$
process at the $\sqrt{s}=14~{\rm TeV}$ LHC. (b) for the $pp \to
W_H^{+} Z_H \to W^+A_H H A_H+X$ process at the $\sqrt{s}=8~{\rm
TeV}$ LHC. (c) for the $pp \to W_H^{-} Z_H \to W^-A_H H A_H+X$
process at the $\sqrt{s}=14~{\rm TeV}$ LHC. (d) for the $pp \to
W_H^{-} Z_H \to W^-A_H H A_H+X$ process at the $\sqrt{s}=8~{\rm
TeV}$ LHC. }
\end{center}
\end{figure}
\begin{figure}[htbp]
\begin{center}
\includegraphics[width=0.45\textwidth]{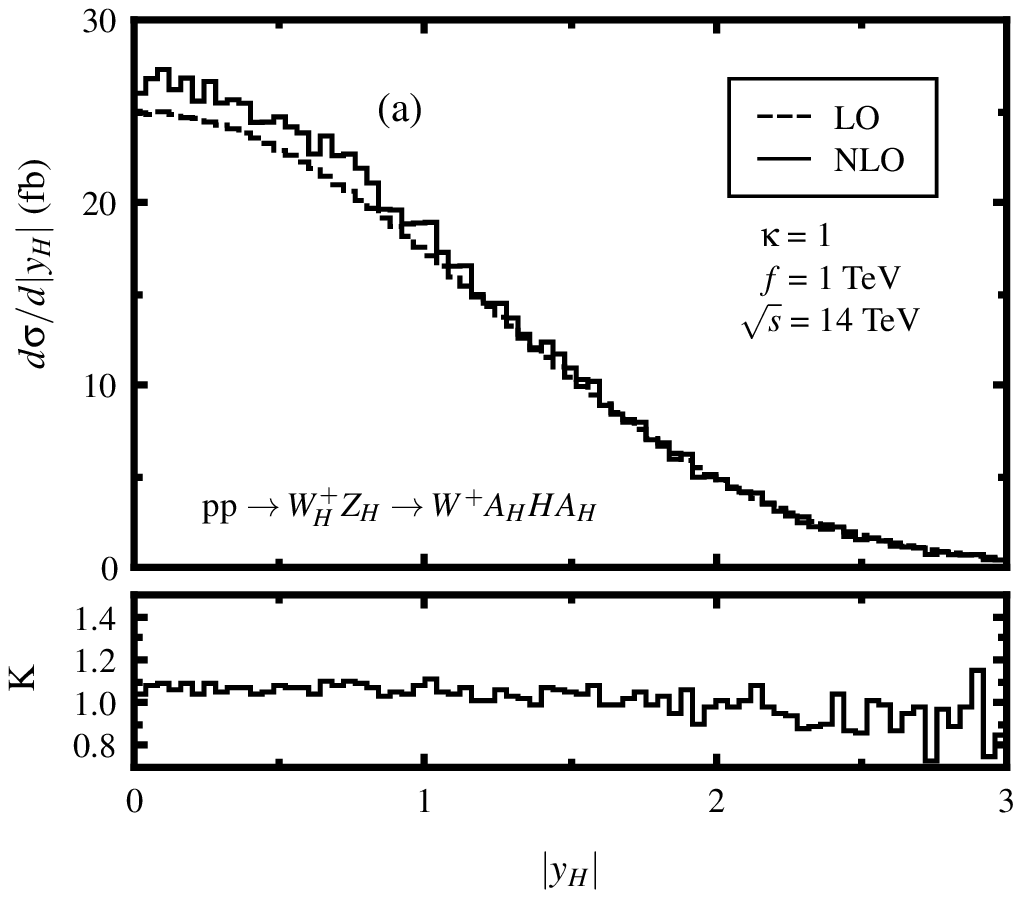}
\includegraphics[width=0.45\textwidth]{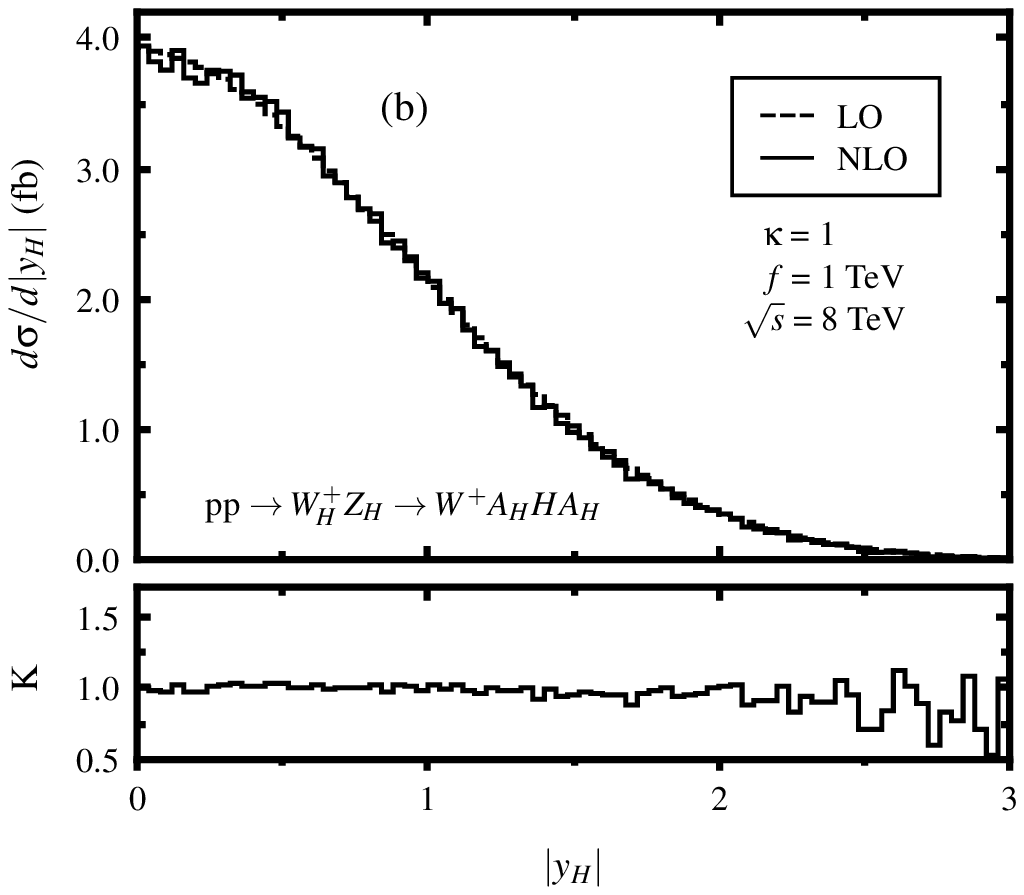}
\\~~  \\
\includegraphics[width=0.45\textwidth]{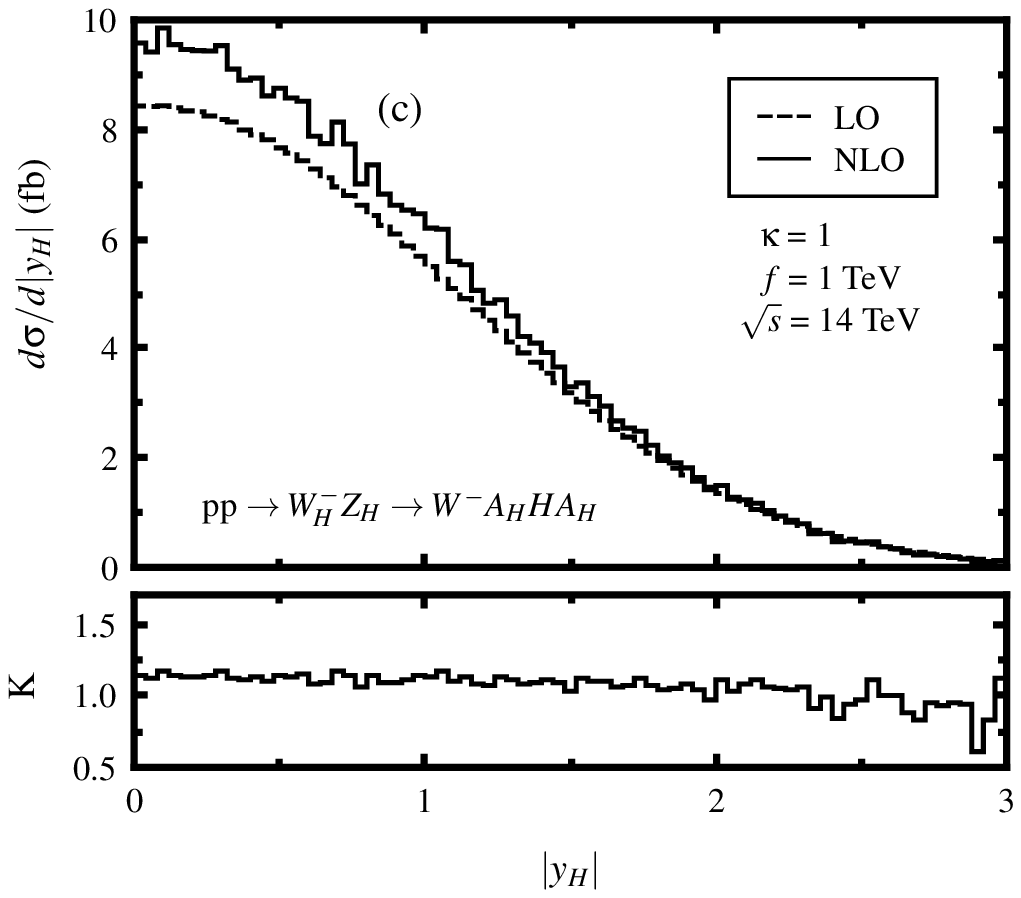}
\includegraphics[width=0.45\textwidth]{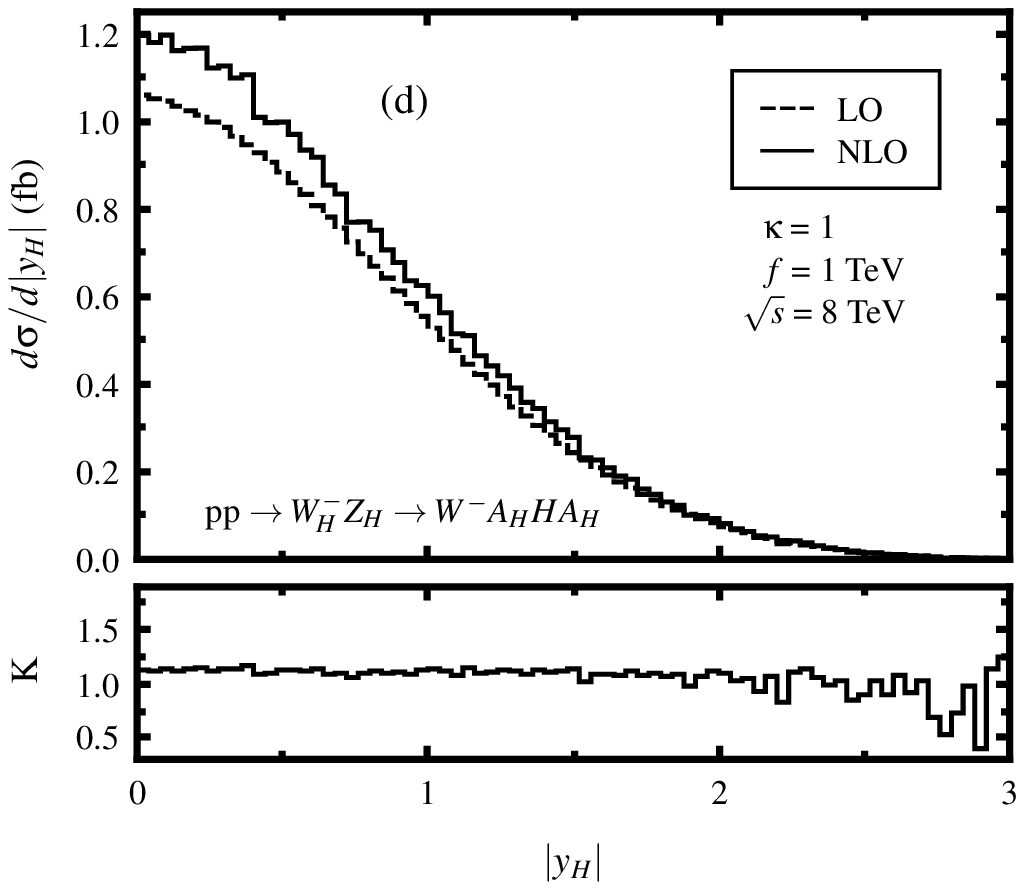}
\caption{\label{fig13} The LO, QCD NLO corrected corrected rapidity
distributions of final Higgs boson, $|y_{H}|$, and the corresponding
$K$-factors in scheme (II) for the $pp \to W_H^{\pm} Z_H \to W^\pm
A_H H A_H+X$ processes by taking $f=1~{\rm TeV}$ and $\kappa=1$. (a)
for the $pp \to W_H^{+} Z_H \to W^+A_H H A_H+X$ process at the
$\sqrt{s}=14~{\rm TeV}$ LHC. (b) for the $pp \to W_H^{+} Z_H \to
W^+A_H H A_H+X$ process at the $\sqrt{s}=8~{\rm TeV}$ LHC. (c) for
the $pp \to W_H^{-} Z_H \to W^-A_H H A_H+X$ process at the
$\sqrt{s}=14~{\rm TeV}$ LHC. (d) for the $pp \to W_H^{-} Z_H \to
W^-A_H H A_H+X$ process at the $\sqrt{s}=8~{\rm TeV}$ LHC.  }
\end{center}
\end{figure}
\begin{figure}[htbp]
\begin{center}
\includegraphics[width=0.45\textwidth]{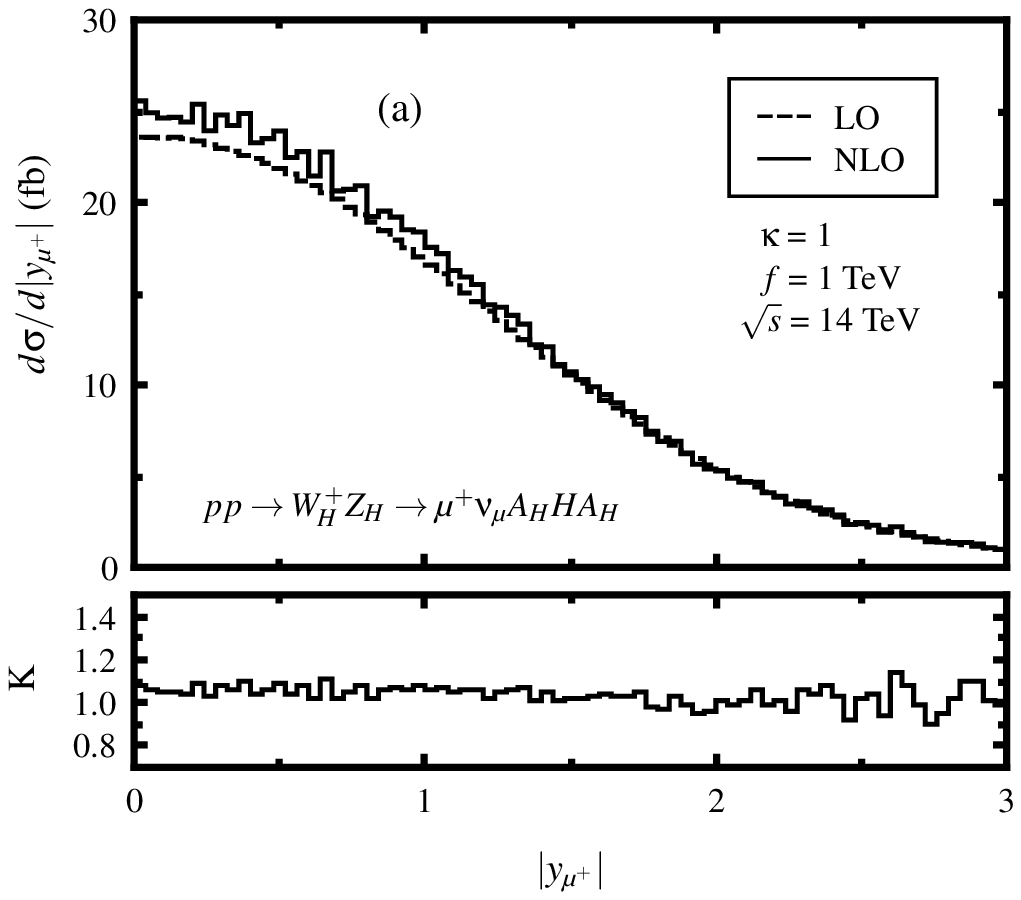}
\includegraphics[width=0.45\textwidth]{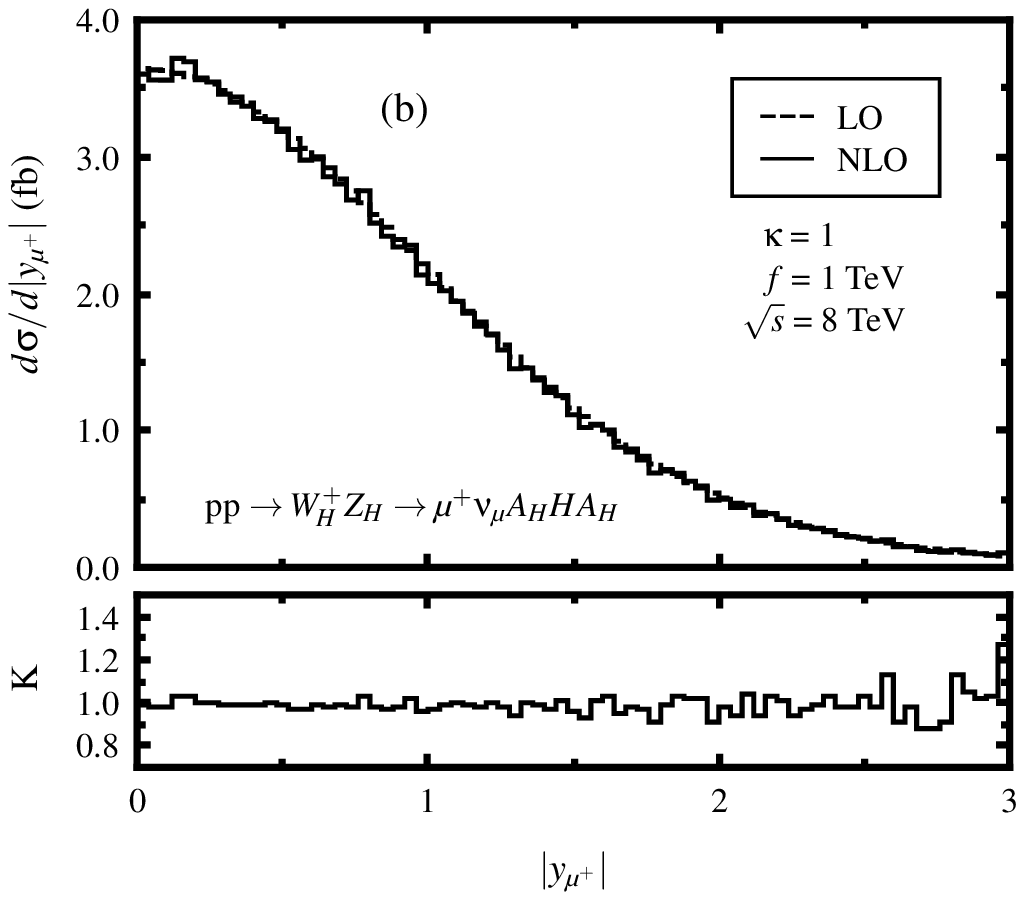}
\\~~  \\
\includegraphics[width=0.45\textwidth]{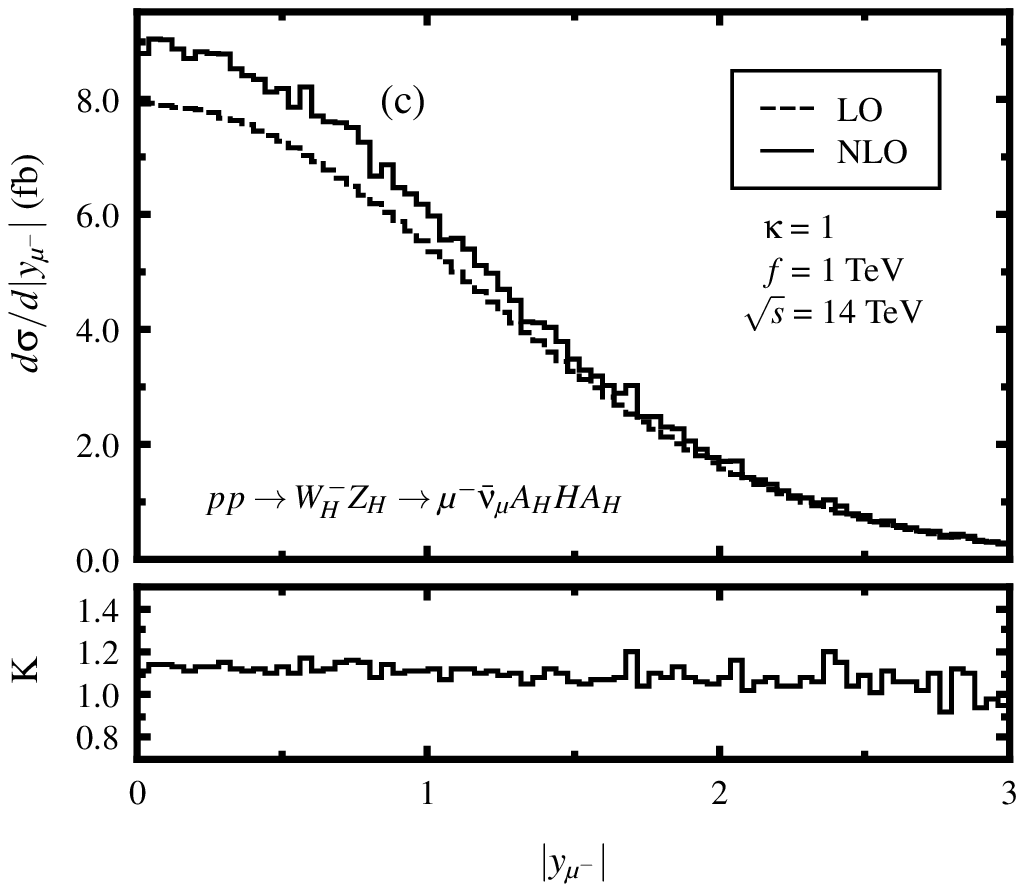}
\includegraphics[width=0.45\textwidth]{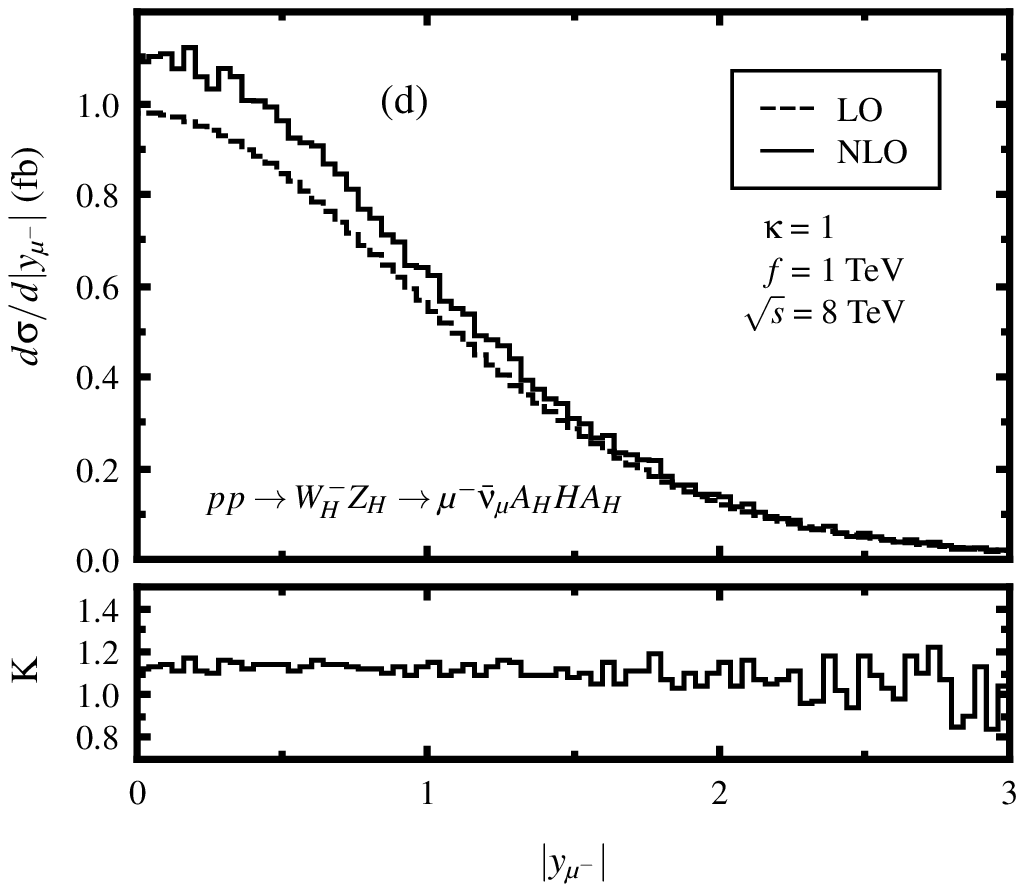}
\caption{\label{fig14} The LO, QCD NLO corrected rapidity
distributions of final muon lepton and the corresponding $K$-factors
in scheme (II) for the $pp \to W_H^{\pm}Z_H \to \mu^{\pm}
\stackrel{(-)}{\nu_{\mu}} A_H H A_H+X~$ processes by taking
$f=1~{\rm TeV}$ and $\kappa=1$. (a) for the \ppfinalbp~ process at
the $\sqrt{s}=14~{\rm TeV}$ LHC. (b) for the \ppfinalbp~ process at
the $\sqrt{s}=8~{\rm TeV}$ LHC. (c) for the \ppfinalbm~ process at
the $\sqrt{s}=14~{\rm TeV}$ LHC. (d) for the \ppfinalbm~ process at
the $\sqrt{s}=8~{\rm TeV}$ LHC.  }
\end{center}
\end{figure}

To show how the ${\cal O}(\alpha_s)$ contributions correct the LO
differential cross sections at the future and early LHC, we depict
the LO, QCD NLO corrected rapidity distributions of final $W$-boson
and Higgs boson ($|y_{W}|$ and $|y_H|$) for the $ W_H^{\pm} Z_H$
production processes in Figs.\ref{fig12}(a,b,c,d) and
Figs.\ref{fig13}(a,b,c,d), respectively. The $W^+$ and Higgs boson
rapidity distributions of the $pp \to W_H^{+} Z_H \to W^+A_H H
A_H+X$ process at the future and early LHC are depicted in
Figs.\ref{fig12}(a,b) and Figs.\ref{fig13}(a,b), respectively.
Figs.\ref{fig12}(c,d) and Figs.\ref{fig13}(c,d) provide the
$|y_{W^-}|$ and $|y_H|$ distributions of the $pp \to W_H^{-} Z_H \to
W^-A_H H A_H+X$ process, which offer the comparisons with
Figs.\ref{fig12}(a,b) and Figs.\ref{fig13}(a,b) correspondingly. The
rapidity distributions of the final $\mu$-lepton ($|y_{\mu}|$) at
the LO and QCD NLO are presented in Figs.\ref{fig14}(a,b,c,d). The
$y_{\mu^+}$ distributions for the \ppfinalbp~ process at the
$\sqrt{s}=14~{\rm TeV}$ and $\sqrt{s}=8~{\rm TeV}$ LHC are plotted
in Figs.\ref{fig14}(a,b) separately, while the $y_{\mu^-}$
distributions for the \ppfinalbm~ process at the $\sqrt{s}=14~{\rm
TeV}$ and $\sqrt{s}=8~{\rm TeV}$ LHC are shown in
Figs.\ref{fig14}(c,d) respectively. All these figures are obtained
by taking the LHT parameters $f=1~{\rm TeV}$, $\kappa=1$ and
adopting the event selection scheme (II). The corresponding
$K$-factors are also plotted in each nether plot of
Figs.\ref{fig12}, Figs.\ref{fig13} and Figs.\ref{fig14}. We can see
from all these figures that the QCD NLO corrections do not make
shape change in the rapidity distributions.

\vskip 5mm
\section{Summary}
\par
We present the calculations of the $W_H Z_H$ production at the CERN
LHC up to the QCD NLO in the littlest Higgs model with $T$ parity.
The dependence of the cross section on the
factorization/renormalization scale are investigated theoretically,
and the rapidity and transverse momentum distributions of final
decay products at both LO and NLO are presented. For the purpose of
providing reliable predictions on the \ppwz process at the LHC, we
adopt two event selection schemes in considering the QCD NLO
corrections for comparison. By using the inclusive scheme the
perturbative convergence could be destroyed, while we can keep the
convergence of the perturbative QCD description and get moderate QCD
NLO corrections to the production rate with evidently reduced scale
uncertainty by adopting the PROSPINO subtraction scheme and setting
$\mu_F=\mu_R$. With this scheme the QCD NLO correction enhances the
LO cross section, and the corresponding $K$-factor for the $W_H^+
Z_H$ production process at the future (early) LHC varies in the
range of $1.01 \sim 1.10$ $(1.00 \sim 1.08)$ when $f$ goes from
$400~{\rm GeV}$ to $1.5~{\rm TeV}$ ($1~{\rm TeV}$), while the
$K$-factor for the $W_H^- Z_H$ production process at the future
(early) LHC varies in the range of $1.11 \sim 1.13$ ($1.11 \sim
1.12$) in the same region of $f$.

\vskip 5mm
\par
\noindent{\large\bf Acknowledgments:} This work was supported in
part by the National Natural Science Foundation of China (Grants No. 11075150, No. 11005101, No. 11275190)
and the Fundamental Research Funds for the Central Universities (Grant No. WK2030040024).

\vskip 5mm
\section{Appendix }
\par
We list the Feynman rules for the coupling vertices in the LHT
related to this work in Table \ref{tabA-1}
\cite{Hubisz:2004ft,cpyuan:2006ph,THan,KPan}, where
$P_{L,R}=\frac{1}{2}(1\mp \gamma_5)$ and $v= v_{SM}$.
\begin{table}[h]
\tiny
\begin{center}
\begin{tabular}{|c|l||c|l|}
\hline
Vertex & ~~~~~~~~Feynman rule & Vertex & ~~~~~~~~~Feynman rule \\
\hline
&&& \\
$W^{+\mu}(k_1) W_{H}^{-\nu}(k_2) Z_{H}^\rho(k_3)$ & $i\frac{e}{s_w}
\left[ g^{\mu \nu} (k_1-k_2)^\rho+\right.$ & $\bar q_-^\alpha
q_-^\beta G_\mu^a$ &
$ig_s(T^a)_{\alpha \beta}\gamma_\mu$\\
&&& \\
 & $\left. g^{\nu \rho} (k_2-k_3)^\mu+g^{\rho \mu} (k_3-k_1)^\nu \right]$ &
 & $$\\
&&& \\
$W^+_{H\mu} \bar U_{i-} D_{j}(i,j=1,2)$ & $i\frac{g}{\sqrt{2}}
\gamma_\mu P_L(V_{Hd})_{ij}$ &
$W^-_{H\mu} \bar D_{i-} U_{j}(i,j=1,2)$ & $i\frac{g}{\sqrt{2}} \gamma_\mu P_L(V_{Hu})_{ij}$ \\
&&& \\
$Z_{H\mu}\bar U_{i-} U_{j}(i,j=1,2)$ & $i\left(\frac{g
c_H}{2}-\frac{g' s_H}{10}\right)\gamma_\mu P_L (V_{Hu})_{ij}$ &
$Z_{H\mu}\bar D_{i-} D_{j}(i,j=1,2)$ & $i\left(-\frac{g
c_H}{2}-\frac{g' s_H}{10}\right)\gamma_\mu P_L (V_{Hd})_{ij}$ \\
&&& \\
\hline
\end{tabular}
\caption{\label{tabA-1} The related LHT Feynman rules used in this
work,  $U_i=u,c$, $D_i=d,s$, $U_{i-}=u_-,c_-$ and
$D_{i-}=d_-,s_-$. $i$ is the generation index. }
\end{center}
\end{table}


\vskip 5mm
\begin{thebibliography}{99}
\bibitem{LHP}
  R. Barbieri and A. Strumia, Phys. Lett. {\bf B462}, 144 (1999).

\bibitem{LittleHiggs}
  N. Arkani-Hamed, A. G. Cohen and H. Georgi, Phys. Lett. {\bf B513}, 232 (2001);
  M. Schmaltz and D. Tucker-Smith, Annu. Rev. Nucl. Part. Sci. {\bf 55}, 229 (2005);
  M. Perelstein, Prog. Part. Nucl. Phys. {\bf 58}, 247 (2007), and references therein.

\bibitem{s1}
  S. L. Glashow, Nucl. Phys. {\bf 22}, 579 (1961);
  S. Weinberg, Phys. Rev. Lett. {\bf 19}, 1264 (1967);
  A. Salam, in Proc. 8th Nobel Symposium Stockholm, 1968, edited by N. Svartholm
  (Almquist \& Wiksells, Stockholm, 1968), p.367;
  H. D. Politzer, Phys. Rep. {\bf 14}, 129 (1974).

\bibitem{s2}
  P. W. Higgs, Phys. Lett. {\bf 12}, 132 (1964); Phys. Rev. Lett. {\bf 13},
  508 (1964); Phys. Rev. {\bf 145}, 1156 (1966);
  F. Englert and R. Brout, Phys. Rev. Lett. {\bf 13}, 321 (1964);
  G. S. Guralnik, C. R. Hagen and T. W. B. Kibble, Phys. Rev. Lett. {\bf 13}, 585 (1964);
  T. W. B. Kibble, Phys. Rev. {\bf 155}, 1554 (1967).

\bibitem{Limit-f}
  C. Csaki, J. Hubisz, G. D. Kribs, P. Meade and J. Terning, Phys. Rev. {\bf D67}, 115002 (2003).

\bibitem{Wmass}
  ATLAS Collaboration, Phys. Lett. {\bf B 705}, 28 (2011).

\bibitem{Zmass}
  D. Olivito (ATLAS collaboration), at the Meeting of the Division 
of Particles and Fields of the American Physical Society (DPF), 
August 9-13, 2011, Brown University, Providence, Rhode Island 
(to be published), arXiv:1109.0934.

\bibitem{Low:2004xc}
  I. Low, JHEP {\bf 10} (2004) 067.

\bibitem{Hubisz:2004ft}
  J. Hubisz and P. Meade, Phys. Rev. {\bf D71}, 035016 (2005).

\bibitem{Hubisz:2005tx}
  J. Hubisz, P. Meade, A. Noble and M. Perelstein, JHEP {\bf 01} (2006) 135.

\bibitem{Barbieri:2000gf}
  R. Barbieri and A. Strumia, ``The `LEP paradox'", arXiv:hep-ph/0007265.

\bibitem{Cheng:2003ju}
  H. C. Cheng and I. Low, JHEP {\bf 09} (2003) 051; {\bf 08} (2004) 061.

\bibitem{Asano}
  A. Birkedal, A. Noble, M. Perelstein and A. Spray, Phys. Rev. {\bf D74}, 035002 (2006);
  M. Asano, S. Matsumoto, N. Okada and Y. Okada, Phys. Rev. {\bf D75}, 063506 (2007).

\bibitem{YanH}
  R.-Y. Zhang, H. Yan, W.-G. Ma, S.-M. Wang, L. Guo and L. Han,
  Phys. Rev. {\bf D85}, 015017 (2012).

\bibitem{DuSM}
  S.-M. Du, L. Guo, W. Liu, W.-G. Ma and R.-Y. Zhang,
  Phys. Rev. {\bf D86}, 054027 (2012).

\bibitem{cpyuan:2006ph}
  A. Belyaev, C.-R. Chen, K. Tobe and C.-P. Yuan,
  Phys. Rev. {\bf D74}, 115020 (2006).

\bibitem{qhcao}
  Q.-H. Cao and C.-R. Chen, Phys. Rev. {\bf D76}, 075007 (2007).

\bibitem{LH5}
  I. Low, W. Skiba and D.Smith, Phys. Rev. {\bf D66},
  072001 (2002).

\bibitem{FeynArts}
  T. Hahn, Comput. Phys. Commun. {\bf 140}, 418 (2001).

\bibitem{FormCalc}
  T. Hahn and M. Perez-Victoria, Comput. Phys. Commun. {\bf 118}, 153 (1999).

\bibitem{tcpss}
  B. W. Harris and J. F. Owens, Phys. Rev. {\bf D65}, 094032 (2002).

\bibitem{KLN}
   T. Kinoshita, J. Math. Phys. (N.Y.) {\bf 3}, 650 (1962); T. D. Lee and M.
   Nauenberg, Phys. Rev. {\bf 133}, B1549 (1964).

\bibitem{PROSPINO}
  W. Beenakker, R. H\"opker, M. Spira and P. M. Zerwas, Nucl. Phys. {\bf B492},
  51 (1997); W. Beenakker, M. Klasen, M. Kr\"amer, T. Plehn, M. Spira and P. M. Zerwas, Phys.
  Rev. Lett. {\bf 83}, 3780 (1999).

\bibitem{on-shell}
   T. Plehn and C. Weydert, Proc. Sci., CHARGED2010 (2010) 026 [arXiv:1012.3761];
   T. Binoth, D. Goncalves-Netto, D. Lopez-Val, K. Mawatari, T. Plehn and I.
   Wigmore, Phys. Rev. {\bf D84}, 075005 (2011).

\bibitem{Blanke:2007ckm}
  M. Blanke, A. J. Buras, A. Poschenrieder, S. Recksiegel, C.
  Tarantino, S. Uhlig and A. Weiler, JHEP {\bf 01} (2007) 066.

\bibitem{databook}
  K. Nakamura {\it et al.}, J. Phys. {\bf G37}, 075021 (2010).

\bibitem{dipole}
   S. Dittmaier, Nucl. Phys. {\bf B565}, 69 (2000);
   M. Roth, Ph.D. thesis, ETH Z\"urich [Institution Report No. 13363, (1999)].

\bibitem{THan}
   T. Han, H. E. Logan, B. McElrath and L. T. Wang, Phys. Rev. {\bf
   D67}, 095004 (2003).

\bibitem{KPan}
   K. Pan, R.-Y. Zhang, W.-G. Ma, H. Sun, L. Han and Y. Jiang,
   Phys. Rev. {\bf D76}, 015012 (2007).

\end{thebibliography}
\end{document}